\documentclass[aps,prx,twocolumn,preprintnumbers,amsmath,amssymb,floatfix]{revtex4-2}

\usepackage{amsmath,amssymb,amsthm}
\usepackage{dcolumn}
\usepackage{bm}
\usepackage{mathtools}
\usepackage{xcolor}
\usepackage{graphicx}
\usepackage{placeins}
\usepackage{tikz}
\usetikzlibrary{arrows.meta,positioning,calc}
\usepackage{hyperref}
\graphicspath{{figures/}{./}}

\definecolor{fisherblue}{RGB}{36,90,160}
\definecolor{riskred}{RGB}{190,60,45}
\definecolor{reservegreen}{RGB}{35,135,90}
\definecolor{softgray}{RGB}{245,245,245}
\hypersetup{
  colorlinks=true,
  linkcolor=fisherblue,
  citecolor=fisherblue,
  urlcolor=fisherblue
}

\newtheorem{theorem}{Theorem}

\newtheorem{corollary}{Corollary}
\newtheorem{proposition}{Proposition}

\newcommand{\CVaR}{\mathrm{CVaR}}

\newcommand{\E}{\mathbb{E}}
\newcommand{\Prob}{\mathbb{P}}
\newcommand{\Tr}{\mathrm{Tr}}

\newcommand{\ITC}{\mathcal{I}_{\alpha}^{\mathrm{TC}}}
\newcommand{\Ttwostar}{T_2^\star}

\newcommand{\keybox}[1]{%
  \par\smallskip\noindent
  \fcolorbox{fisherblue}{softgray}{%
    \parbox{0.93\columnwidth}{\small #1}}%
  \par\smallskip}

\begin{document}

\title{Fisher Glasses: Tail-Certified Quantum Metrology in Quenched Environments}

\author{El Mustapha Mansouri}
\affiliation{School of Engineering, Institute of Science Tokyo, Yokohama, Kanagawa, 226-8501, Japan}

\author{Keigo Arai}
\email[Corresponding author: ]{arai.k.835f@m.isct.ac.jp}
\affiliation{School of Engineering, Institute of Science Tokyo, Yokohama, Kanagawa, 226-8501, Japan}
\date{\today}

\begin{abstract}
Quantum metrological advantage is certified by averaged Fisher responses:
contrast, susceptibility, or quantum Fisher information (QFI). This fails in
quenched sensors, where slow environmental variables freeze within a session
but vary between repetitions: shallow nitrogen-vacancy (NV) centers,
superconducting qubits with slow two-level fluctuators, and semiconductor spin
qubits in drifting charge noise. They sample session-resolved Fisher geometries,
not an averaged channel. Certification conditions on the latent session,
projects nuisance directions, inverts to attainable loss, then tail-certifies;
this inverse upper-tail loss defines quenched tail-certified information. A
no-go theorem: no averaged Fisher data determine this certificate; ensembles
sharing averaged Fisher matrix, QFI, and projected information have finite or
zero certified precision. A Fisher-zero integrability transition governs
collapse: the inverse-loss tail exponent $\beta$ sets the boundary, with
nonintegrable certified loss for $\beta\leq1$, even when annealed information is
large or scaling. The certified
quantum resource is response transverse to latent disorder, not raw
amplification sharing its generator; universal design laws: safe windows,
nondegenerate portfolios, Fisher reserves, action separation, Fisher-cut
criteria. A shallow-NV Ramsey tournament shows average-QFI optimization is
tail-catastrophic, whereas tail-certified designs recover nearly three orders of
magnitude in certified information at equal shot budget and latent ensemble.
These non-self-averaging phases are Fisher glasses, governed by Fisher-zero
rare-event statistics.
\end{abstract}

\maketitle

% ============================================================
\section{Introduction: average QFI certifies the wrong object}
\label{sec:intro}
% ============================================================

What does a quantum Fisher information (QFI) certificate certify? In the
standard theory it certifies the local distinguishability of a fixed or
self-averaged channel: a larger phase shift, susceptibility, contrast, or QFI
is read as a larger guaranteed precision~\cite{Helstrom1976,Braunstein1994,
Paris2009}. Entangled probes, critical sensors, collective spin states,
Greenberger--Horne--Zeilinger (GHZ) protocols, and hybrid solid-state sensors
are all built to amplify the imprint of a target parameter on a quantum
state~\cite{Giovannetti2004,Giovannetti2006,Giovannetti2011,Degen2017,Pezze2018}, and
realistic noise is known to modify or destroy ideal quantum-metrological
scaling~\cite{Huelga1997,Escher2011,DemkowiczDobrzanski2012,Smirne2016}. All of
this presumes that the experiment actually samples the channel whose QFI is
computed.

Many solid-state quantum sensors violate this presumption at the session
scale. A shallow nitrogen-vacancy (NV) center near sparse surface spins, a
superconducting qubit coupled to individual two-level fluctuators, or a
semiconductor spin qubit in a slowly varying charge environment sees a single
frozen disorder configuration throughout one calibration-and-sensing session,
while nominally identical sessions sample different
configurations~\cite{Romach2015,Paladino2014,Galperin2006,Bergli2009,
Bylander2011,Muller2019,Yoneda2018,Curtis2025}. The experiment is therefore
quenched, not annealed: the data in a session are generated by a conditional
channel selected by a hidden environmental configuration, not by the averaged
channel used in a design calculation. The shallow NV makes the stakes tangible
(and becomes a quantitative design tournament in Sec.~\ref{sec:nv-platform}).
The positions and couplings of its surface spins are frozen while the device is
calibrated and run; most freezes are benign, but a rare one places a
bath-coherence node inside the chosen interrogation window and erases the usable
information for that entire session. A designer who optimizes the average
response can fund, build, and deploy a sensor that is certified precise on paper
yet operationally useless on exactly the sessions that decide the measurement.

This is the failure of an average taken over the wrong object. Averaging the
Fisher response over the latent ensemble is like judging a river safe to cross
by its mean depth: the average is reassuring and the arithmetic is exact, but
one never crosses at the average. One fords the particular bed one is given, and
a single deep stretch (one session in which the usable information
collapses) is fatal whatever the benign mean promised. A quantum
sensor likewise faces one frozen environment per session and must survive the
specific configuration it is dealt, not the ensemble average.

Average QFI therefore answers the wrong operational question. It certifies the
distinguishability of an averaged channel, a fiction that no single run
ever realizes. A quenched experiment instead asks a sharper question: in the one
frozen session that actually occurs, once the slow environmental directions are
set aside as nuisance directions, what loss is attainable? This distinction
matters because precision is governed by inverse information, so a small set of
sessions in which the usable information nearly vanishes can dominate the
certified loss even when the average response is large. Certifying the average
therefore does not certify the experiment one runs.

\begin{figure*}[t]
\centering
\includegraphics[width=\textwidth]{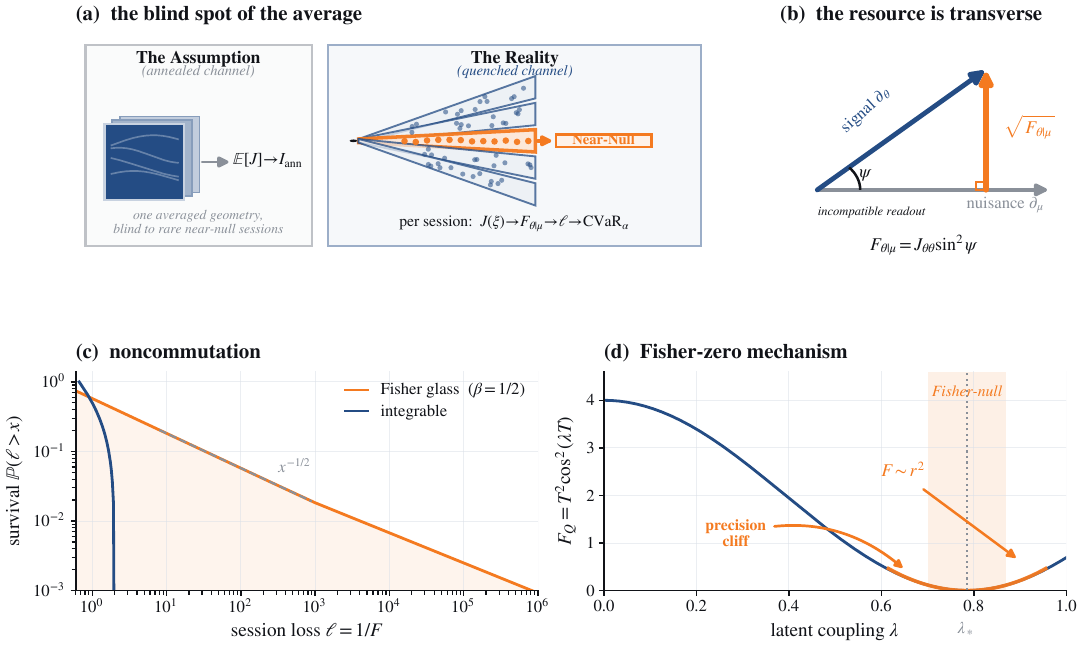}
\caption{\textbf{Quenched Fisher geometry: a visual abstract.}
(a) The averaged-channel paradigm certifies one geometry,
$\E[J]\to$ average QFI, whereas the quenched, non-self-averaging experiment
samples a
distribution of session-resolved geometries $J(\xi)$, each fed through the
noncommuting pipeline $J\to F_{\theta|\mu}\to\ell\to\CVaR_\alpha$ of
Eq.~\eqref{eq:intro-pipeline}, so rare near-null sessions ($F_\perp\approx0$)
hidden by the average dominate the loss. (b) The certified resource is the
response transverse to the latent nuisance,
$F_{\theta|\mu}=J_{\theta\theta}\sin^2\psi$; for a vector target the obstruction
is measurement incompatibility, quantified by the Holevo loss of
Sec.~\ref{sec:transverse}. (c) Two latent ensembles with the same mean
$\E[F]=1$ but different inverse-loss tails, one integrable (finite
certificate) and one Fisher-glass ($\beta=1/2$, $\mathcal{I}^{\rm TC}_\alpha=0$),
that no averaged Fisher quantity can distinguish. (d) Microscopic mechanism: a
simple Fisher zero $F_Q=T^2\cos^2(\lambda T)\sim r^2$ near a node gives
$\Prob(F_Q^{-1}>x)\sim x^{-1/2}$, the generic nonintegrable Fisher-glass tail.}
\label{fig:paradigm}
\end{figure*}

The repair is an order-of-operations change, as described by
Eq.~\eqref{eq:intro-pipeline}. One must first condition on the
frozen session, then project the slow nuisance directions, then invert the
remaining information to the attainable loss, and only then certify its upper
tail over future sessions,
\begin{equation}
  \xi
  \;\longrightarrow\;
  J(\xi)
  \;\longrightarrow\;
  F_{\theta|\mu}(\xi)
  \;\longrightarrow\;
  \ell(\xi)
  \;\longrightarrow\;
  \CVaR_\alpha[\ell].
  \label{eq:intro-pipeline}
\end{equation}
We call the inverse of this upper-tail loss the tail-certified information. When
this certificate vanishes while the averaged response remains large, we call the
regime a \emph{Fisher glass}. The formal definitions are given in
Sec.~\ref{sec:experiment}. The central claim is that this order cannot be
reconstructed after averaging: once the latent-session distribution is compressed
to an averaged Fisher matrix or averaged QFI, the tail of the inverse loss has
already been erased, and no later certificate can recover it.
Figure~\ref{fig:paradigm} contrasts the two paradigms.

We develop the certification theory of this quenched Fisher geometry and
establish four results. First (Sec.~\ref{sec:no-average}), averaged Fisher
data are insufficient: two latent ensembles can share the same averaged Fisher
response yet have different tail-certified precision, including the case of
finite precision for one and zero for the other, and the tail-certified risk is
in general not even a renormalized Fisher matrix. Second
(Sec.~\ref{sec:transition}), the certificate is controlled by the tail of the
inverse loss, so a simple Fisher zero already produces a Fisher glass, while
higher-codimension failures and nondegenerate portfolios restore integrability.
Third (Secs.~\ref{sec:transverse}--\ref{sec:scaling}), the certified quantum
resource is not raw amplification but the component of the signal response
transverse to the latent-disorder directions, with a Holevo extension for vector
targets. Fourth (Secs.~\ref{sec:repair}--\ref{sec:nv-platform}), these principles
become concrete design laws (safe interrogation windows, nondegenerate
portfolios, configuration-wise Fisher reserves), and a shallow-NV Ramsey
tournament shows that the average-QFI optimum can be tail-catastrophic while
tail-certified designs recover orders of magnitude in certified information.

\paragraph*{Relation to neighboring frameworks.}
The distinguishing feature is the order forced by frozen latent sessions in
Eq.~\eqref{eq:intro-pipeline}: CVaR, nuisance projection, and Fisher zeros
become a precision certificate only when composed in this order. Decoherence-limited QFI
theory~\cite{Huelga1997,Escher2011,DemkowiczDobrzanski2012,Smirne2016},
fixed-model nuisance
metrology~\cite{SuzukiYangHayashi2020,Ragy2016,Szczykulska2016},
Bayesian~\cite{ZhangSuzuki2025,AlbarelliBranfordRubio2026} and
minimax~\cite{OuyangShettellMarkham2021,VishnupriyaHarikrishnanPal2026}
metrology, and compound/random-channel and finite-sample or outage-style
analyses~\cite{MeyerKhatriFrancaEisertFaist2025} each average, fix, or
worst-case the model at an earlier stage; quenched Fisher-glass certification
instead treats the projected attainable loss itself as a random variable over
frozen session labels and certifies its upper tail. The framework-by-framework
comparison is given in the Supplemental Methods
(Table~\ref{tab:risk-comparison}). Throughout, ``advantage'' refers to the
certified local information gain of a quantum-enhanced protocol relative to its
chosen benchmark, and this comparison must be made with the quenched
tail-certified loss rather than averaged QFI.

% ============================================================
\section{Quenched metrological experiment}
\label{sec:experiment}
% ============================================================

We consider nominally identical estimation sessions performed in a
non-self-averaging environment. At the start of session
$i$, a latent environmental configuration $\xi_i$ is sampled from a
device/environment law $\pi_\xi$ and held fixed through that session's
calibration and sensing shots. The target parameter $\theta$ is a
deterministic local parameter common to the benchmarked sessions and is
\emph{not} assigned a prior; the risk is over future session labels, not over
$\theta$. Conditional on $\xi_i$, data are generated by a local model
$\rho^\xi_{\theta,\mu}$, where $\mu$ denotes slow nuisance coordinates that
affect the record but are not reported targets. Calibration may estimate
session-specific contrast, curvature, or a reduced description of $\xi_i$; it
conditions the estimator on the current session but does not replace the
experiment by an annealed channel. Throughout, $\xi\sim\pi_\xi$ is the
session-fixed configuration, $J(\xi)$ is the conditional Fisher (or QFI)
matrix, scalar and directional Fisher informations are denoted by $F$ with
subscripts, and $\ell_{\mathcal P}(\xi)$ is the conditional estimation loss of
protocol $\mathcal P$.

\subsection{Notation and the quenched certification pipeline}
\label{sec:notation}

The quenched certification is assembled in the fixed, noncommuting order of
Eq.~\eqref{eq:intro-pipeline}: condition on the frozen session $\xi$, project the
nuisance directions, invert to a loss, and only then tail-certify. These
operations do not commute because averaging before inversion fundamentally
changes the operational precision guarantee. Conditioning
fixes $J(\xi)$; projecting the slow coordinate $\mu$ leaves the usable scalar
information $F_{\theta|\mu}(\xi)$, written $F_\perp(\xi)$ in the general projected
case; and inverting gives the session loss $\ell(\xi)=1/F_{\theta|\mu}(\xi)$. A
short scope summary closes this section, with the full assumptions in
Table~\ref{tab:scope-assumptions} (Sec.~\ref{sec:conclusion}).

The averaged response defines the annealed descriptor
\begin{equation}
  I_{\rm ann}=\E_\xi[F_\perp(\xi)],
  \label{eq:intro-annealed}
\end{equation}
with $\E_\xi\equiv\E_{\xi\sim\pi_\xi}$ the latent-session average; the raw
average QFI $\E_\xi[J_{\theta\theta}(\xi)]$ is even less conservative, since
$J_{\theta\theta}\geq F_\perp$ pointwise (nuisance projection only removes
information). We use ``averaged Fisher descriptor'' broadly for the QFI
$J_{\rm ann}=J(\bar\rho_\theta)$ of the annealed channel
$\bar\rho_\theta=\E_\xi[\rho^\xi_{\theta,\mu}]$ and for averages of conditional
Fisher responses such as $\E_\xi[J_{\theta\theta}(\xi)]$ or
$\E_\xi[F_\perp(\xi)]$; none of these determines the quenched inverse-loss tail.

The operational precision certificate is instead defined by the inverse
upper-tail risk of the conditional loss,
\begin{equation}
  \ITC(\mathcal P)
  :=
  \left[\CVaR_\alpha\!\left(\ell_{\mathcal P}(\xi)\right)\right]^{-1},
  \qquad 0\leq\alpha<1.
  \label{eq:tc-info}
\end{equation}
For scalar or directional targets
$\ell_{\mathcal P}(\xi)=F^{-1}_{\theta|\mu,\mathcal P}(\xi)$; for vector
quantum targets the reported target is $B\theta$ and $\ell_{\mathcal P}$ is the
nuisance-projected Holevo loss $C^H_{B\theta|\mu}(W_B;\xi)$, with positive weight
$W_B$ on the reported target components, of Sec.~\ref{sec:transverse}. The
conditional value-at-risk $\CVaR_\alpha(X)$ is the mean of the worst $1-\alpha$
fraction of the loss distribution, a coherent upper-tail risk
measure~\cite{Artzner1999,Rockafellar2000,Rockafellar2002}; its dual form and the
CVaR--QCRB bound are derived in Appendix~\ref{app:cvar}. When the CVaR in
Eq.~\eqref{eq:tc-info} is infinite, the certified information is zero. A protocol
can therefore have large or scaling $I_{\rm ann}$ yet
$\mathcal{I}^{\rm TC}_{\alpha}=0$. This establishes the defining signature of a
Fisher glass: quantum enhancement survives in the annealed Fisher geometry while
disappearing as an operational quenched resource. The collapse is reached by two
mechanisms invisible to the averaged geometry. Either the amplified response
aligns with a latent nuisance direction, so signal and environment become
locally indistinguishable; or rare configurations make the usable information
collapse. The law $\pi_\xi$ may come from a physical
latent-environment model or from the empirical distribution of repeated
calibration sessions; for a single slowly drifting device the same certificate
is a block-sampled time-series certificate when the drift is stationary or
slowly mixing, and otherwise a finite-catalog empirical certificate.

\keybox{\textbf{Core objects and notation.} A frozen session label
$\xi\sim\pi_\xi$ fixes the conditional Fisher matrix $J(\xi)$; nuisance
projection gives the usable scalar information $F_\perp(\xi)=F_{\theta|\mu}(\xi)$,
the loss $\ell(\xi)=1/F_\perp(\xi)$, and, for an estimation protocol $\mathcal P$
with loss $\ell_{\mathcal P}$, the certificate
$\ITC(\mathcal P)=[\CVaR_\alpha(\ell_{\mathcal P}(\xi))]^{-1}$ of
Eq.~\eqref{eq:tc-info}.
Every later object specializes this pipeline.}

The pointwise quantum Cram\'er--Rao bound (QCRB) is the operational bridge from
information to loss. For a locally unbiased scalar estimator with conditional
mean-square loss $R(\xi;\theta)$ and $n$ independent shots, monotonicity of
CVaR applied to the pointwise bound $R(\xi;\theta)\geq1/[nF_Q(\theta,\xi)]$, with
$F_Q(\theta,\xi)$ the session-conditional scalar QFI,
gives
\begin{equation}
  \CVaR_\alpha(R(\xi;\theta))
  \geq
  \CVaR_\alpha\!\left(\frac{1}{nF_Q(\theta,\xi)}\right).
  \label{eq:cvar-qcrb}
\end{equation}
Thus a protocol with large average QFI has zero tail-certified information
whenever $\CVaR_\alpha(1/F_Q)=+\infty$. When the inverse-loss survival function
obeys
\begin{equation}
  \Prob(\ell_{\mathcal P}>x)\sim A_\beta x^{-\beta},
  \qquad 0<A_\beta<\infty ,
  \label{eq:tail-cert-beta}
\end{equation}
finite certified information is equivalent to $\beta>1$. The confidence level
$\alpha$ is a reporting choice; the tail-integrability boundary $\beta=1$ is not,
because if the inverse-loss mean is nonintegrable then every finite-level
upper-tail CVaR is infinite in the ideal latent distribution. The multiparameter,
directional, and Holevo versions of Eq.~\eqref{eq:cvar-qcrb}, together with the
CVaR preliminaries, are collected in the Supplemental Methods
(Appendix~\ref{app:cvar}). The remaining sections identify the operational loss
$\ell_{\mathcal P}$ and the physical mechanisms that set $\beta$.

\begingroup
\footnotesize
\setlength{\tabcolsep}{4pt}
\renewcommand{\arraystretch}{1.25}
\newcommand{\sca}[1]{\parbox[t]{0.47\columnwidth}{\raggedright #1}}
\begin{center}
\textbf{Scope of the main statements} (full assumptions in
Table~\ref{tab:scope-assumptions}).
\smallskip

\begin{tabular}{@{}ll@{}}
\hline\hline
Result & \sca{Main assumption}\\
\hline
CVaR--QCRB certificate & \sca{session-fixed latent $\xi$; local regularity}\\
Geometric Fisher-zero exponent & \sca{smooth latent density; local zero $F\sim r^{2m}$ (vanishing order $m$)}\\
Portfolio law $x^{-K/2}$ ($K$ arms) & \sca{independent, zero-nondegenerate arms}\\
Shallow-NV tournament & \sca{sparse quasi-static random-telegraph-noise (RTN) model}\\
\hline\hline
\end{tabular}
\end{center}
\endgroup

\FloatBarrier

% ============================================================
\section{No-go theorem for annealed certification}
\label{sec:no-average}
% ============================================================

The conceptual core of the theory is a no-go statement: the quenched
certificate is not a functional of the averaged Fisher geometry. This is
stronger than saying average QFI is loose; it says average QFI discards
information that decides whether the certificate is finite at all. The
following theorem shows that no quantity derived from the averaged Fisher
geometry can serve as an operational precision certificate in a quenched
experiment.

\begin{theorem}[No averaged Fisher certification]
\label{thm:no-average}
There exist latent-session ensembles with the same averaged Fisher matrix
$\overline J=\E_\xi[J(\xi)]$ and the same averaged projected information
$\E_\xi[F_{\theta|\mu}(\xi)]$, but different tail-certified information
\begin{equation}
  \mathcal{I}^{\rm TC}_{\alpha,\theta|\mu}
  =
  \left[\CVaR_\alpha\!\left(F_{\theta|\mu}^{-1}(\xi)\right)\right]^{-1},
  \label{eq:noavg-itc}
\end{equation}
including one with $\mathcal{I}^{\rm TC}_{\alpha,\theta|\mu}>0$ and one with
$\mathcal{I}^{\rm TC}_{\alpha,\theta|\mu}=0$. Consequently, operational
certification is encoded in the quenched inverse-loss statistics rather than in
any averaged Fisher geometry.
\end{theorem}

It suffices to treat the scalar case and embed it as the $\theta$ block of a
larger Fisher matrix. One ensemble has $F(\xi)=1$ deterministically, so
$\CVaR_\alpha(F^{-1})=1$ and the certificate is finite. The other is chosen
with bounded support, density nonzero near $F=0$ so that $F^{-1}$ has a
nonintegrable upper tail, and compensating mass at larger finite $F$ so that
$\E F=1$. The two ensembles share the averaged Fisher matrix and the average
projected information, yet the CVaR of $F^{-1}$ is finite for the first and
infinite for the second. Latent certification therefore depends on the
projected inverse-loss tail, not on the averaged geometry. The explicit
construction is given in Appendix~\ref{app:noncommutation}. Panel
\ref{fig:paradigm}(c) shows the two ensembles; they are indistinguishable to
any averaged Fisher quantity.

\subsection{Breakdown of Fisher geometry}
\label{sec:no-effective}

The failure is not repaired by replacing $J$ with some effective
$J_{\rm eff}$. Consider the tail-certified directional risk for target
direction $w$ (with $J(\xi)^{-1}$ read as the Moore--Penrose inverse
$J(\xi)^{+}$ on singular strata, as in Appendix~\ref{app:cvar}),
\begin{equation}
  R_\alpha(w)
  =
  \CVaR_\alpha\!\left(w^\top J(\xi)^{-1}w\right).
  \label{eq:risk-body}
\end{equation}
By CVaR duality, $R_\alpha$ is a supremum of positive quadratic forms, hence
convex and homogeneous of degree two in $w$, so its unit sublevel set is a
symmetric convex \emph{risk body} (Appendix~\ref{app:cvar}). But it is in
general not quadratic. For two equally likely configurations with
$J_1^{-1}=\operatorname{diag}(1,100)$ and
$J_2^{-1}=\operatorname{diag}(100,1)$ at $\alpha=1/2$,
\begin{equation}
  R_{1/2}(w)=\max\!\left\{w^\top J_1^{-1}w,\ w^\top J_2^{-1}w\right\},
  \label{eq:risk-body-example}
\end{equation}
which violates the parallelogram identity and therefore cannot arise from any
single matrix. In ordinary local metrology a Fisher matrix defines an ellipsoid
of distinguishability; in the quenched tail-certified theory the risk body is a
convex body that is generally not an ellipsoid. Tail certification is thus not
a renormalized QFI but a genuinely different information geometry, which is the
precise sense in which the quenched Fisher geometry is a new object.

% ============================================================
\section{Fisher-glass transition: the Fisher-zero exponent}
\label{sec:transition}
% ============================================================

Theorem~\ref{thm:no-average} says the certificate lives in the inverse-loss
tail; the next result says exactly when that tail is integrable. A
\emph{Fisher zero} is any latent configuration where the scalar, directional,
or nuisance-projected Fisher information vanishes, whether from bath coherence nodes,
channel failure, or loss of transverse identifiability when the signal response
becomes tangent to a latent-disorder direction.

The exponent $\beta$ is the inverse-loss tail index of the quenched experiment.
It is not an averaged Fisher quantity: it is defined only after conditioning on
the frozen session, projecting nuisance directions, and inverting to the
attainable loss. Operationally, $\beta$ is the certification exponent:
$\beta>1$ gives a finite tail-certified information, while $\beta\le1$ is the
Fisher-glass regime. Geometrically, Fisher-zero mechanisms set $\beta$ through
codimension and vanishing order; experimentally, $\beta$ is the slope measured
from session-loss tails.

\begin{theorem}[Fisher-glass transition]
\label{thm:fisher-glass}
Let $F_\perp(\xi)$ be the session-resolved projected Fisher information. If, near
a Fisher-null set,
\begin{equation}
  \Prob(F_\perp<\varepsilon)\sim A_\beta\varepsilon^\beta,
  \qquad 0<A_\beta<\infty,
  \label{eq:fperp-small-tail}
\end{equation}
then
\begin{equation}
  \Prob(F_\perp^{-1}>x)\sim A_\beta x^{-\beta},
  \label{eq:fperp-inverse-tail}
\end{equation}
so $\mathcal{I}^{\rm TC}_{\alpha}=0$ for $\beta\leq1$, even when
$I_{\rm ann}=\E_\xi[F_\perp(\xi)]$ is large or scaling. More generally, if the
Fisher-null set is a smooth codimension-$c$ manifold and, in normal distance
$r$, $F_\perp\sim r^{2m}$ and the latent density $p_\xi\sim r^\nu$, then
$\Prob(F_\perp^{-1}>x)\sim A_\eta x^{-\eta}$ with the geometric Fisher-zero
exponent
\begin{equation}
  \eta=\frac{c+\nu}{2m},
  \label{eq:codim-exponent}
\end{equation}
so that $\beta=\eta$ when the tail is produced by a geometric Fisher zero, and
finite certification requires $\eta>1$.
\end{theorem}

\paragraph*{Physics behind the theorem.}
The mechanism is elementary. Let a latent coordinate $r$ measure distance from a
simple Fisher zero. Generically $F_\perp(r)\simeq a_\star r^2$. If sessions sample
$r$ with smooth nonzero density at the zero, then
\begin{equation}
  \Prob(F_\perp<\varepsilon)
  =
  \Prob\!\left(|r|<\sqrt{\varepsilon/a_\star}\right)
  \propto
  \sqrt{\varepsilon},
  \label{eq:simple-zero-small}
\end{equation}
hence $\Prob(F_\perp^{-1}>x)\propto x^{-1/2}$. Because
\begin{equation}
  \E[F_\perp^{-1}]
  =
  \int_0^\infty\Prob(F_\perp^{-1}>x)\,dx
  \label{eq:layercake}
\end{equation}
diverges for tail index $\beta\leq1$, the upper-tail mean loss diverges:
a single simple zero in the latent support is already catastrophic.
\keybox{\textbf{Why an average response is dangerous.} The annealed
descriptor $\E[F]$ is insensitive to the integrability of $1/F$, which is
exactly the quantity the tail certificate measures. Two protocols with the same
$\E[F]$ can sit on opposite sides of the $\beta=1$ boundary.}
This is the whole physical content of the transition: a codimension-one simple
zero ($c=1$, $m=1$, $\nu=0$) gives $\eta=1/2$, while increasing the codimension
of simultaneous failure moves $\eta$ across the tail-integrability boundary
$\eta=1$. The geometric proof is in Appendix~\ref{app:codim-proof}.

These same tail indices double as an experimental diagnostic and preview the
repairs of Sec.~\ref{sec:repair}, as the next proposition records.

\begin{proposition}[Design principles for escaping the Fisher glass]
\label{prop:tail-slope-diagnostics}
Fisher zeros and weakest Fisher cuts set the inverse-loss tail index, while safe
windows, nondegenerate portfolios, and Fisher reserves change it in
directly measurable ways. The signatures are tail indices: $\beta=1/2$ ($x^{-1/2}$)
for a single unsafe zero, $\beta=K/2$ ($x^{-K/2}$) for a zero-nondegenerate
$K$-arm portfolio, and a safe Fisher reserve changes the universality class by
eliminating the algebraic tail altogether.
\end{proposition}

\paragraph*{Fisher-zero universality.}
Many microscopic noise models produce the same certification exponent $\beta$.
In pure-dephasing sensors, real zeros of the bath characteristic function or
Loschmidt amplitude are a microscopic source of Fisher zeros~\cite{Gorin2006,Quan2006,Heyl2013,Heyl2018}. If the coherence
has first real zero $u_1$ and the disorder action $A$ (the evolution time
setting the accrued latent phase $A\lambda$) satisfies $A\Lambda\ge u_1$
(with $\Lambda$ the maximum latent coupling),
neighborhoods of the zero generate the same algebraic inverse-loss tail; for
quasi-static two-state noise $u_1=\pi/2$
(Appendix~\ref{app:loschmidt-proof}). For Ramsey magnetometry at the optimal
working point the conditional QFI is $F_Q(T,\xi)=T^2\mathcal{W}^2(T,\xi)$, with $\mathcal{W}$ the
coherence selected by the latent configuration. In the quasi-static two-state
limit $W_{\rm RTN}=\cos(\lambda T)$, so an unsafe single Ramsey arm has the
generic $x^{-1/2}$ inverse-Fisher tail. With $\gamma_{\rm RTN}$ the
random-telegraph-noise switching rate, the exact safety boundary is
\begin{equation}
  T_{\rm safe}(\Lambda,\gamma_{\rm RTN})
  =
  \frac{\pi/2+\arcsin(\gamma_{\rm RTN}/\Lambda)}
  {\sqrt{\Lambda^2-\gamma_{\rm RTN}^2}}
  \label{eq:tsafe-main}
\end{equation}
(Appendix~\ref{app:rtn}). For uniform couplings and sparse Poisson
fluctuators, $T<T_{\rm safe}$ keeps the Fisher zero outside the latent support,
whereas $T\ge T_{\rm safe}$ gives infinite CVaR of $1/F_Q$ for every
$\alpha<1$. With Gaussian background contrast of envelope time $T_2^\star$ and mean
fluctuator number $N_f$, the conventional average-QFI Ramsey optimum is
$T_{\rm avg}=T_2^\star/\sqrt2+O(N_f)$ (maximizing $T^2e^{-2(T/T_2^\star)^2}$), and in the
tail-catastrophic regime used below the tail-certified (CVaR-optimal)
interrogation time $T_{\rm TC}$ satisfies
\begin{equation}
  T_{\rm TC}<T_{\rm safe}<T_{\rm avg} .
  \label{eq:tcvar-safe-tavg}
\end{equation}
The usual Ramsey design can therefore be average-optimal and tail-catastrophic
at once.

\paragraph*{Operational regulators in finite data.}
Real experiments have finite data, bounded estimators, calibration floors, and
priors, so every empirical CVaR is finite. This finiteness is not a loophole in
the criterion but a regime the theory predicts quantitatively: each regulator
converts the sharp $\beta=1$ transition into a finite precision cap that an
experimentalist can compute in advance from the calibration budget.
\keybox{\textbf{Finite-data regulators are predictive engineering laws.} A hard
loss cap $\ell_{\max}$, additive Fisher floor $\varepsilon$, or finite latent
catalog $N_{\rm eff}$ caps the certified information at a value fixed by the
regulator:
$\mathcal{I}^{\rm TC}_{\alpha,\ell_{\max}}\propto\ell_{\max}^{-(1-\beta)}$,
$\mathcal{I}^{\rm TC}_{\alpha,\varepsilon}\propto\varepsilon^{1-\beta}$, and
$\mathcal{I}^{\rm TC}_{\alpha,N}\asymp A\,N_{\rm eff}^{-(1-\beta)/\beta}$ (with
logarithmic forms at $\beta=1$). Read forward, these regulator-flow laws
predict where a design's certified precision plateaus for given hardware; read
backward, they size the calibration catalog or loss cap needed to reach a
target certificate.}
A certificate that still moves with its regulator is not yet intrinsic:
regulator-independent certification requires either $\beta>1$ or a physical
Fisher reserve with $F_\perp(\xi)\geq F_0>0$. Used deliberately, however, the
regulator flow is itself a quantitative design tool, proved in
Appendix~\ref{app:regulator-flow}.

% ============================================================
\section{The certified quantum resource is transverse}
\label{sec:transverse}
% ============================================================

The quenched certificate changes the meaning of quantum enhancement under
latent disorder. The relevant question is not whether the signal susceptibility
is large, but whether it remains identifiable after slow environmental
directions are projected out.

\subsection{Transverse resource: the physical picture}

Before the matrix statement, the mechanism is worth seeing physically. The
obstruction appears whenever the target and the latent disorder are written
into the probe through a shared generator. This is the situation of trying to
hear one tone over a second, nearly identical tone: turning up the
volume (raw amplification) boosts signal and background together and leaves
them exactly as hard to tell apart. Gain along the shared direction buys no
distinguishability; what is needed is a different direction, a component of the
response that the disorder does not also produce.

Geometrically, collect the slow latent directions into a nuisance tangent space
and picture it as a plane, on which the target imprints a response vector. If
that vector lies in the plane, no amount of gain separates signal from
environment: its length can grow without bound while its projection onto the
orthogonal complement stays zero. Certifiable information is carried only by the
part of the response transverse to the plane, which for a single nuisance
direction is measured by the \emph{Fisher angle} $\psi$ between the signal
response and that direction. As $\psi\to0$ the response folds into the nuisance plane and the
certified information vanishes no matter how large the raw susceptibility. The
Schur-complement algebra of Appendix~\ref{app:transverse-volume} is precisely
the projection that extracts this
transverse component for any number of nuisance directions, and for one it
collapses to the one-line law $F_{\theta|\mu}=J_{\theta\theta}\sin^2\psi$.

\subsection{Scalar nuisance projection}

\begin{theorem}[The quantum resource is transverse]
\label{thm:transverse}
Promote a slow latent coordinate to a session-fixed nuisance parameter $\mu$,
and let $\theta$ be the target. On a regular stratum where the nuisance block is
nonsingular, projecting the nuisance out of the session QFI matrix $J(\xi)$
leaves the nuisance-projected Fisher information $F_{\theta|\mu}$, the Schur
complement of the nuisance block, whose general form and multi-channel geometry
are derived in Appendix~\ref{app:transverse-volume}. For a single nuisance
coordinate it collapses to the one-line law
\begin{equation}
  \boxed{\,F_{\theta|\mu}=J_{\theta\theta}\sin^2\psi\,},
  \label{eq:fisher-angle}
\end{equation}
where the Fisher angle $\psi$ is defined by
$\cos\psi=J_{\theta\mu}/(J_{\theta\theta}J_{\mu\mu})^{1/2}$. The operational
inverse-Fisher loss for estimating $\theta$ while $\mu$ is unknown is
$F_{\theta|\mu}^{-1}$, and the certified scalar information in the presence of
the nuisance is
\begin{equation}
  \mathcal{I}^{\rm TC}_{\alpha,\theta|\mu}
  =
  \left[\CVaR_\alpha\!\left(F_{\theta|\mu}^{-1}\right)\right]^{-1}.
  \label{eq:projected-tail-certified}
\end{equation}
\end{theorem}

The projection must be applied session by session, \emph{before} inversion and
tail certification; otherwise the averaged Fisher geometry hides nonintegrable
rare-session losses. Equation~\eqref{eq:fisher-angle} makes the obstruction
explicit: a divergent scalar QFI does not certify enhancement if the Fisher
angle to the nuisance direction closes. The certified resource is not the raw
susceptibility $J_{\theta\theta}$ but the susceptibility transverse to the
latent-disorder tangent space. The same identifiability principle extends to
multi-channel sensors: if all active channels respond to signal and disorder in
the same ratio, the target is locally indistinguishable from a nuisance
displacement and the projected information vanishes
(Sec.~\ref{sec:repair} and Appendix~\ref{app:transverse-volume})~\cite{Proctor2018,ZhangZhuang2021,Guo2020}.

\keybox{\textbf{Raw amplification that shares a generator with the disorder is
not a tail-certified resource.} Gain applied along the signal--disorder common
direction raises $J_{\theta\theta}$ and the latent-disorder response in the
same ratio, leaving the Fisher angle $\psi$ (and hence the certified
transverse information $J_{\theta\theta}\sin^2\psi$) unchanged. Certified
enhancement requires increasing the response transverse to the disorder, not
the raw susceptibility.}

\subsection{Vector targets and the Holevo loss}

\textit{Measurement incompatibility.}~
The scalar inverse-Fisher loss is only the compatible-measurement limit of the
theory. Before the formalism, the geometric picture extends to several targets
at once. Where a scalar target was a single response vector poking out of the
nuisance plane at angle $\psi$, a vector target is a rigid bundle of response
vectors (a multidimensional shape) protruding from that same plane, and two
obstructions now compound. Each component still loses whatever part of its
response lies in the nuisance plane; but in addition the components cannot all
be read out optimally at once, because the single measurement setting that best
resolves one component is not the setting that best resolves another, and these
optimal settings are mutually exclusive. This obstruction is the
non-commutativity of the per-target optimal measurements; unlike a classical
shortfall it is not removed by more signal, and persists at any amplification. The
certified loss is then the minimum total unavoidable blur over the bundle,
weighted by the importance $W_B$ assigned to each component, namely the volume of the
protruding shape that survives both the nuisance projection and this measurement
trade-off. The Holevo loss below is the
quantitative form of exactly this picture, and raw amplification that merely
enlarges the shape inside the nuisance plane does not reduce the blur. In a
genuinely quantum multiparameter problem one specifies the
reported target as
\begin{equation}
  \vartheta=B\theta,\qquad B\in\mathbb R^{d_B\times p},
  \label{eq:reported-target-main}
\end{equation}
with $B$ full row rank and positive weight $W_B\succ0$, and replaces
$F_{\theta|\mu}^{-1}$ by the attainable nuisance-projected Holevo loss
$C^H_{B\theta|\mu}(W_B;\xi)$. The certificate is then
\begin{equation}
  \mathcal{I}^{\rm TC,H}_{\alpha,B|\mu}(W_B)
  =
  \left[
  \CVaR_\alpha\!\left(C^H_{B\theta|\mu}(W_B;\xi)\right)
  \right]^{-1}.
  \label{eq:main-holevo-target-certificate}
\end{equation}
Tail certification is applied to this session-resolved attainable loss, not to
the symmetric-logarithmic-derivative (SLD) Fisher matrix, so the vector
Fisher-glass criterion is the tail integrability of $C^H_{B\theta|\mu}$. A
minimal pure-qubit example in Appendix~\ref{app:holevo} shows that the target
map $B$ distinguishes an unidentifiable full target from an identifiable
reported component: when an unknown nuisance shifts one quadrature, the full
two-parameter target becomes unidentifiable
($C^H_{\theta|\mu}=\infty$) while a reported transverse direction remains
finite. Scalar, directional, and full-vector estimation are the cases
$d_B=1$, $B=w^\top$, and $B=I_p$~\cite{Holevo1982,
AlbarelliFrielRobertsonDatta2019,YangChiribellaHayashi2019,SuzukiYangHayashi2020}.

\paragraph*{Quantum attainable loss and measurement incompatibility.}
The session-resolved loss is a \emph{quantum attainable loss} (the QCRB for
scalar targets and the Holevo loss for incompatible multiparameter targets), so
measurement incompatibility enters through
$C^H_{B\theta|\mu}\ge\operatorname{tr}[W_B\,BJ_{\theta|\mu}^{+}B^\top]$, and the
certified object is a quenched quantum information geometry, not classical
post-processing of an externally given Fisher number. The interpretation of
quantum enhancement mechanisms changes accordingly: GHZ, collective,
critical~\cite{Zanardi2006,Zanardi2008,FrerotRoscilde2018,Rams2018,Garbe2020},
and error-corrected~\cite{Kessler2014,Dur2014,Arrad2014,Sekatski2017}
amplification certify only through their Fisher-transverse component $g_\perp$,
not through raw gain $g_s$ (Sec.~\ref{sec:scaling}).

% ============================================================
\section{Action separation and tail-safe quantum enhancement}
\label{sec:scaling}
% ============================================================

\begin{figure*}[t]
\centering
\includegraphics[width=\textwidth]{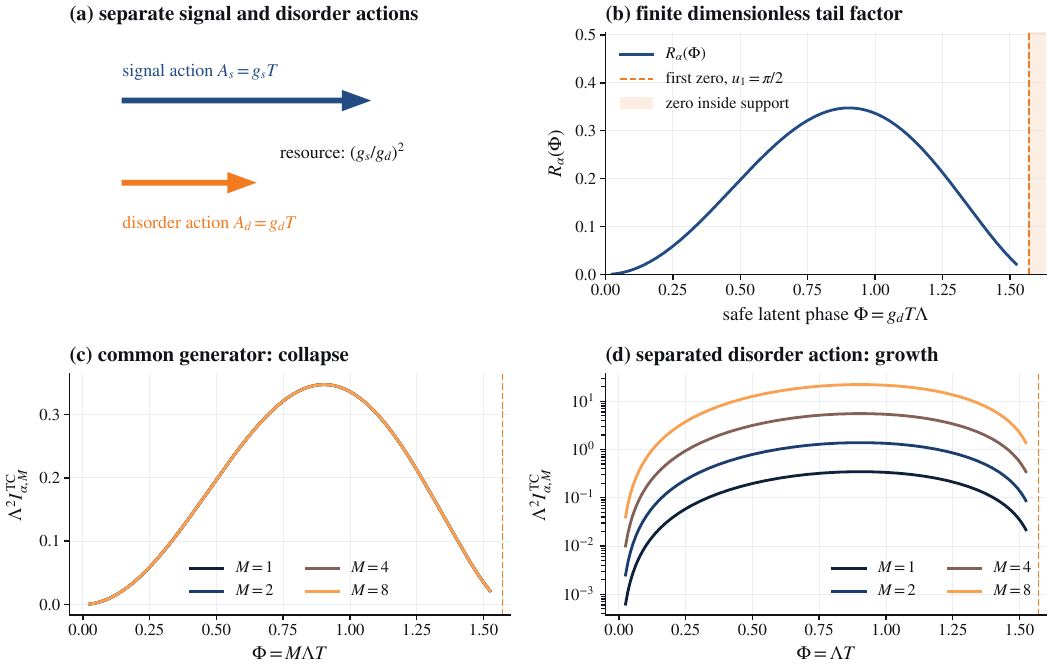}
\caption{\textbf{Signal--disorder action separation.}
Tail-certified amplification is controlled by the ratio of transverse signal
action to disorder action. (a) A protocol writes signal and latent disorder
into the probe with generally different action gains $g_s$ and $g_d$; the
fixed-safe-phase resource is $(g_s/g_d)^2$, which becomes $(g_\perp/g_d)^2$
after nuisance projection. (b) Dimensionless tail factor
$\mathcal{R}_\alpha(\Phi)$, with $\Phi$ the dimensionless safe latent phase, for a
two-state Loschmidt response, with first zero
at $u_1=\pi/2$; the certified information vanishes once the latent support
reaches the zero. (c) Common-generator amplification, $g_s=g_d=M$ and
$\Phi=M\Lambda T$, collapses $\Lambda^2\mathcal{I}^{\rm TC}_{\alpha,M}$ for
different $M$. (d) Separated disorder action, $g_s=M$ and $g_d=1$, gives
$\Lambda^2\mathcal{I}^{\rm TC}_{\alpha,M}=M^2\mathcal{R}_\alpha(\Phi)$,
restoring scalable per-shot tail-certified gain. In the full multiparameter
setting $g_s$ is replaced by the nuisance-projected transverse gain
$g_\perp=g_s\sin\psi$.}
\label{fig:action-separation}
\end{figure*}

Applying the transverse projection of Theorem~\ref{thm:transverse} to a family
of amplified protocols turns the certificate into a scaling law: amplification
helps only when it raises the identifiable transverse response faster than the
latent-disorder response. Let $g_s$ be the raw signal action gain and $g_d$ the
disorder action gain, and define the transverse gain $g_\perp$ by
\begin{equation}
  g_\perp^2T^2\equiv F_{\theta|\mu}^{\rm amp},
  \qquad
  g_\perp^2=g_s^2\sin^2\psi
  \label{eq:gperp-definition}
\end{equation}
up to smooth contrast and dimensionless safe-phase factors. At fixed tail-safe
latent phase, the per-shot tail-certified information is controlled by
\begin{equation}
  \boxed{\,
  \mathcal{I}^{\rm TC}_{\alpha,\mathrm{shot}}
  \propto
  \left(\frac{g_\perp}{g_d}\right)^2,\,}
  \label{eq:transverse-action-law}
\end{equation}
not by $g_s^2$. This is the transverse signal--disorder action-separation law
(Fig.~\ref{fig:action-separation}, Appendix~\ref{app:action-separation-proof}).

The interpretation is immediate. If signal and latent disorder share a common
generator, $g_s=g_d=M$, then whenever a latent Fisher zero is present the
tail-safe interrogation time must shrink as $T=O(1/M)$ to stay before the zero.
This cancels the nominal $M^2$ per-shot gain: the optimized certified per-shot
information is $O(1)$, up to contrast factors (Fig.~\ref{fig:action-separation}(c));
separating the disorder action from the signal instead restores the $M^2$ gain
(Fig.~\ref{fig:action-separation}(d)). Common-generator GHZ,
collective, and critical amplification are therefore \emph{not} tail-certified
resources by themselves; scalable certified enhancement requires
$g_\perp/g_d\to\infty$, a Fisher reserve, or physical removal of the zero. This
is the metrological counterpart of the known impossibility of Heisenberg-limited
sensing with simply entangled probes in a generic open
environment~\cite{Huelga1997,Escher2011,DemkowiczDobrzanski2012}. For a
critical-sensor family of size $L$ with $g_\perp\sim L^{y_\perp}$ and
$g_d\sim L^{y_d}$, the scaling
$\mathcal{I}^{\rm TC}_{\alpha,\mathrm{shot}}(L)\sim L^{2(y_\perp-y_d)}$~\cite{Rams2011,Harris1974}, together with network, echo, and DC specializations, is
given in Appendix~\ref{app:transverse-volume}.

% ============================================================
\section{Tail engineering}
\label{sec:repair}
% ============================================================

Because the certificate is set by the inverse-loss tail, the designs are
geometric: they reshape that tail before inversion. There are two kinds.
\emph{Intrinsic} repairs change the conditional Fisher geometry $F_\perp(\xi)$
and can produce regulator-independent certification; \emph{extrinsic}
regularizers only cap or floor the reported loss and obey the regulator-flow
laws of Sec.~\ref{sec:transition}. The intrinsic repairs realize the three
slopes of Proposition~\ref{prop:tail-slope-diagnostics}. Each repair below is
stated as a transformation of the inverse-loss tail and paired with its
consequence on the shallow-NV apparatus of Sec.~\ref{sec:nv-platform}, whose
tournament certifies all four designs at a common shot budget and latent
ensemble against the average-QFI baseline.

\paragraph*{Safe window.} Choose interrogation times before the first latent
Fisher zero. In the single-arm Ramsey model this is $T<T_{\rm safe}$, and the
optimum is the CVaR-optimal time inside that window. On the NV magnetometer
this single relocation of the working point (from the average-QFI time, which
sits beyond the boundary $T_{\rm safe}$, to a CVaR-optimal time inside the safe
window) lifts the certified information by more than three orders of magnitude
(Sec.~\ref{sec:nv-platform}).

\begin{figure*}[tp]
\centering
\includegraphics[width=\textwidth]{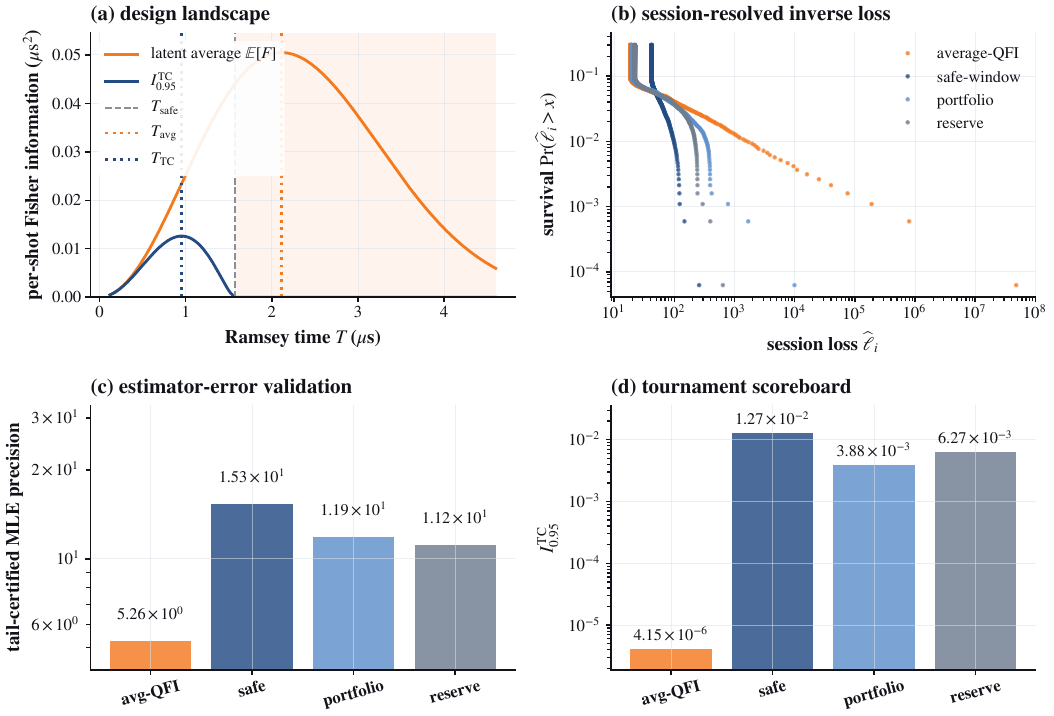}
\caption{\textbf{Annealed winner, quenched loser.}
A platform-level shallow-NV design tournament compares four Ramsey protocols
under the same shot budget and latent session ensemble. (a) The standard
average-QFI rule selects $T_{\rm avg}>T_{\rm safe}$, crossing into the
Fisher-glass region, while the tail-certified optimizer stays at
$T_{\rm TC}<T_{\rm safe}$. (b) Session-resolved inverse-loss tails expose the
mechanism: the average-QFI arm has a much heavier bad-session tail, while
portfolio and reserve designs suppress the far tail. (c) With finite
calibration shots, finite sensing shots, and an explicit MLE, the plotted
tail-certified MLE precision $1/\CVaR_{0.95}(\mathrm{MSE}_{\rm MLE})$ tracks the
predicted ranking. (d) Scoreboard: safe-window, portfolio, and reserve designs
recover nearly three orders of magnitude in $\mathcal{I}^{\rm TC}_{0.95}$
relative to the average-QFI design. This protocol-ranking reversal is the
operational signature of the Fisher-glass regime.}
\label{fig:nv-platform-demo}
\end{figure*}

\paragraph*{Nondegenerate portfolios.} Use $K$ active interrogation times with
positive shot allocations. If the arms are zero-nondegenerate (no latent
configuration kills all arms simultaneously at codimension one), then
$\Prob(\ell_{\rm port}>x)=x^{-K/2+o(1)}$. Hence $K=1$ is catastrophic, $K=2$ is
marginal in the regular case, and $K\geq3$ gives finite CVaR
(Appendix~\ref{app:portfolio-proof}). This threshold is concrete rather than
abstract: a three-arm NV portfolio raises the inverse-loss tail index across the
$\beta=1$ boundary and certifies almost three orders of magnitude above the
average-QFI design at the identical shot budget
(Sec.~\ref{sec:nv-platform}, Fig.~\ref{fig:nv-platform-demo}(d)).

\paragraph*{Fisher reserve.} Unlike the previous two strategies, Fisher reserves
eliminate the Fisher-zero tail altogether rather than merely changing its
exponent. Add a certified safe reference channel with
positive shot allocation. If $F_{\rm tot}(\xi)\geq F_0(\xi)>0$ for
every finite latent configuration, the algebraic Fisher-zero tail is removed;
in the sparse Poisson model the remaining tail is super-polynomial
(Appendix~\ref{app:reserve-proof}). On the optical table this inequality is the
literal addition of a short, zero-free anchor arm: the NV reserve design of
Sec.~\ref{sec:nv-platform} bolts a safe reference channel onto the same
apparatus, certifying roughly three orders of magnitude above average
QFI, while still retaining the long, nominally optimal arm.

\paragraph*{Control and encoding.} Control, echo, decoherence-free design~\cite{Hahn1950,Viola1999,Biercuk2009,deLange2010}, and
error-corrected sensing restore certification when they reduce the
latent-disorder action, rotate the signal response away from the disorder
tangent space, or add a configuration-wise reserve. Same-axis DC sensing is
more constrained, because control that cancels quasi-static latent phase also
reduces signal area, so universal scalar tail safety there requires portfolios,
reserves, or signal--disorder separation (Appendices~\ref{app:echo} and
\ref{app:dc-bound}).

These rules generalize beyond Ramsey arms through a single architecture law. In
a multi-channel sensor one writes the conditional Fisher geometry as a positive
sum of channel contributions, $J_{\theta|\mu}(\xi)=\sum_e g_e(\xi)\,a_ea_e^\top$
with $a_e$ the response direction of channel $e$,
and a \emph{target-identifying Fisher cut} is a set of channels whose
simultaneous failure makes the reported target absorbable into the nuisance
span. In complex sensing architectures, Fisher zeros are no longer isolated
points but propagate through sensing networks, so the relevant object is the
weakest Fisher cut.

\begin{theorem}[Minimum-cut law for Fisher glasses]
\label{thm:fisher-cut}
For a positive-channel decomposition of the conditional Fisher matrix with
independent channel weights $g_e$ satisfying
$\Prob(g_e<\varepsilon)=\varepsilon^{\eta_e+o(1)}$, the nuisance-projected
inverse-loss exponent is set by the weakest target-identifying Fisher cut,
\begin{equation}
  \Prob(F_{\theta|\mu}^{-1}>x)=x^{-\beta_{\theta|\mu}+o(1)},
  \qquad
  \beta_{\theta|\mu}
  =
  \min_{C\in\mathcal C_{\theta|\mu}}\sum_{e\in C}\eta_e ,
  \label{eq:fisher-cut-main}
\end{equation}
where $\mathcal C_{\theta|\mu}$ are the cuts after which every remaining
channel's target response is absorbable into the nuisance span. Finite
tail-certified information requires $\beta_{\theta|\mu}>1$.
\end{theorem}

\begin{figure*}[t]
\centering
\begin{minipage}[b]{0.32\textwidth}
\centering
\includegraphics[width=\linewidth]{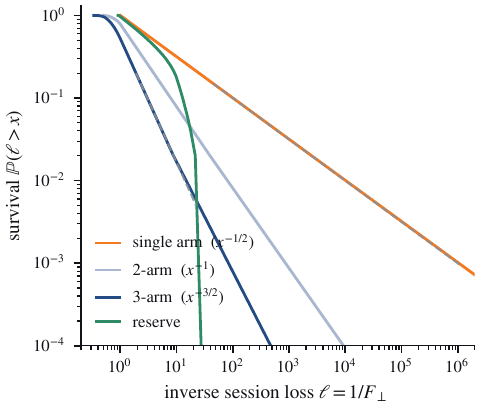}
\end{minipage}\hfill
\begin{minipage}[b]{0.65\textwidth}
\centering
\includegraphics[width=\linewidth]{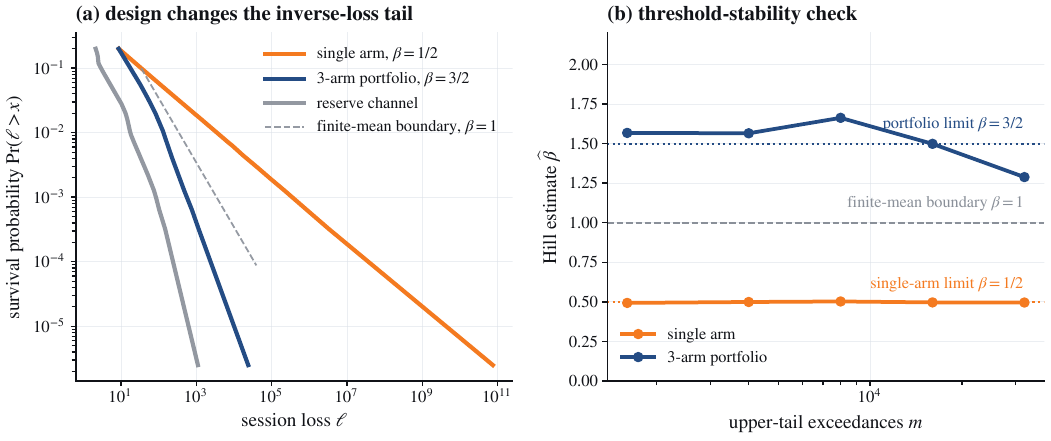}
\end{minipage}
\caption{\textbf{Tail diagnostics from session losses.}
\emph{Left:} survival functions of the inverse session loss $\ell=1/F_\perp$
for the four design classes: a single unsafe arm follows the nonintegrable
$x^{-1/2}$ law (below the finite-mean boundary $\beta=1$), a two-arm portfolio
is marginal ($x^{-1}$), a zero-nondegenerate three-arm portfolio steepens to
$x^{-3/2}$, and a configuration-wise Fisher reserve removes the algebraic zero
tail; the measured slope is the inverse-loss tail index $\beta$. \emph{Right:} experimental
certification from calibrated session losses: (a) survival curves
distinguishing unsafe, portfolio, and reserve designs, and (b) Hill upper-tail
estimates of $\beta$ giving a finite-sample stability check against the $\beta=1$
boundary.}
\label{fig:fisher-glass-transition}
\label{fig:experimental-certification}
\end{figure*}

Theorem~\ref{thm:fisher-cut} generalizes the Fisher-zero exponent from isolated
zeros to arbitrary sensing architectures. The architecture, not the averaged
Fisher matrix, decides the bad-tail
exponent: for $K$ generic channels and $q$ nuisance directions
with identical simple-zero statistics $\eta_e=1/2$, the smallest cut has size
$K-q$ (the $q$ nuisance directions absorb the target response of at most $q$
channels, so the remaining $K-q$ must fail together), hence
$\beta_{\theta|\mu}=(K-q)/2$ and finite certification requires
$K\geq q+3$. The scalar case $q=0$ recovers the portfolio threshold $K\geq3$;
one nuisance direction raises it to $K\geq4$. The proof and its
graph-Laplacian (Fisher-percolation) specialization are in
Appendices~\ref{app:fisher-cut-proof} and \ref{app:transverse-volume}.
Figure~\ref{fig:fisher-glass-transition}~(left) shows the four design classes
as distinct survival slopes.

% ============================================================
\section{Shallow-NV Ramsey design tournament}
\label{sec:nv-platform}
% ============================================================

We close with a design tournament in which the standard design rule selects the
wrong protocol. A shallow nitrogen-vacancy (NV) center is used as a Ramsey
magnetometer in the presence of sparse, slow surface
fluctuators~\cite{Taylor2008,Romach2015,Mamin2013,Staudacher2013,Casola2018,
Rovny2024}. This is not an additional assumption of the theory: each session
samples a fixed local environment while nominally identical sessions sample
different surface-fluctuator configurations, so the NV center is a
platform-level instantiation of the conditional Fisher geometry developed
above. The simulation gives all quantities in physical units, includes finite
readout contrast, and implements an explicit measurement and
maximum-likelihood estimator (MLE), not only an information bound. The
parameters in Table~\ref{tab:nv-platform-parameters} are deliberately
representative: they stress-test the theory in a zero-crossing regime where the
Fisher-null mechanism is active, rather than forecasting one particular device.

\begin{table*}[t]
\centering
\small
\begin{ruledtabular}
\begin{tabular}{llll}
Quantity & Value used & Calibration source & Certificate role\\
\hline
$\gamma_e$ & $0.176\,{\rm rad}\,\mu{\rm s}^{-1}\,\mu{\rm T}^{-1}$ & spin resonance & $B_{\rm sig}\mapsto\theta=\gamma_e B_{\rm sig}$\\
$c_{\rm ro}$ & $0.18$ & optically detected magnetic resonance (ODMR)/readout & readout contrast amplitude\\
$T_2^\ast$ & $3.0\,\mu{\rm s}$ & Ramsey envelope & background contrast\\
$\Lambda$ & $1.0\,{\rm rad}\,\mu{\rm s}^{-1}$ & noise spectroscopy & $T_{\rm safe}=\pi/(2\Lambda)$\\
$\gamma_{\rm RTN}$ & $0$ & quasi-static limit & separates RTN rate from $\gamma_e$\\
$N_f$ & $0.10$ & session contrast statistics & bad-session probability\\
$N_{\rm sess}$ & $32000$ & experimental repetitions & latent-environment samples\\
$n$ & $2000$ & sensing shot budget & estimator variance\\
$R_{\rm err}$ & $8$ & simulated sensing repeats & Fig.~\ref{fig:nv-platform-demo}(c) MLE-error validation\\
$R_{\rm rob}$ & $16$ & simulated sensing repeats & supplemental MLE robustness check\\
$n_{\rm cal}$ & $4000$ & calibration block & curvature uncertainty\\
$\theta_0$ & $0$ & chosen working point & local Ramsey slope\\
seed & $20260430$ & supplementary code & base random seed\\
\end{tabular}
\end{ruledtabular}
\caption{Platform-level parameters for the shallow-NV Ramsey simulation. The
values define a representative stress-test point rather than a device-specific
fit. Supplemental sensitivity sweeps vary the fluctuator density,
dimensionless coupling range, CVaR level, readout contrast, calibration shots,
and wall-clock/dead-time budget around this point; supplemental robustness
checks test finite-session convergence, seed-to-seed variation, tail-exponent
uncertainty, and direct MLE-error validation. At the baseline $\alpha=0.95$
and $N_{\rm sess}=32000$, the empirical CVaR is set by $1600$ tail sessions.
The estimator-error validation in Fig.~\ref{fig:nv-platform-demo}(c) uses
$R_{\rm err}=8$ sensing repeats, while the supplemental direct MLE robustness
check uses $R_{\rm rob}=16$ repeats per calibrated latent session. In an experiment
these inputs are obtained from ODMR, Ramsey, echo, and noise-spectroscopy
calibration.}
\label{tab:nv-platform-parameters}
\end{table*}

The model parameters are listed in Table~\ref{tab:nv-platform-parameters}: $N_f$
is the mean number of active surface fluctuators, $\Lambda$ the latent-coupling
cutoff, $T_2^\star$ the Gaussian Ramsey envelope time, $c_{\rm ro}$ the readout
contrast, and $\theta=\gamma_e B_{\rm sig}$ the target phase per unit time. For
session $i$, a fixed latent configuration is drawn as
\begin{equation}
  \xi_i=\{\lambda_{ij}\}_{j=1}^{N_i},
  \quad
  N_i\sim{\rm Poisson}(N_f),
  \quad
  \lambda_{ij}\sim p_\Lambda,
  \label{eq:nv-latent-config}
\end{equation}
with $p_\Lambda$ uniform on $[0,\Lambda]$. The Ramsey contrast at interrogation
time $T$ is
\begin{equation}
  C_i(T)=\exp[-(T/T_2^\ast)^2]\prod_{j=1}^{N_i}\cos(\lambda_{ij}T),
  \label{eq:nv-session-contrast}
\end{equation}
in the quasi-static two-state limit, and the photon-count readout is the
Bernoulli channel
\begin{equation}
  p_i(1|\theta,T,\varphi)
  =
  \tfrac12\!\left[1+c_{\rm ro}C_i(T)\cos(\theta T+\varphi)\right],
  \label{eq:nv-readout-model}
\end{equation}
with $\theta=\gamma_e B_{\rm sig}$ and optimal sensing phase $\varphi=\pi/2$.
We compare four protocols at the same total shot budget: the \emph{average-QFI}
design uses the time $T_{\rm avg}$ maximizing the latent average of
$F_Q(T,\xi)$; the \emph{safe-window} design uses $T_{\rm TC}<T_{\rm safe}$
maximizing the empirical certificate; the \emph{portfolio} uses three times
$\{1.85,2.20,2.65\}\,\mu$s; and the \emph{reserve} uses a safe anchor plus a
long arm, $\{1.00\,\mu\mathrm{s},T_{\rm avg}\}$. We run the full six-step
tail-certification protocol on the apparatus; the average-QFI design and the
three tail-certified designs pass through identical machinery and differ only in
their interrogation schedule.

\keybox{\textbf{Certification protocol.}
(i)~draw the frozen latent sessions;
(ii)~estimate the conditional Fisher and loss from a calibration block;
(iii)~tail-certify via the empirical CVaR;
(iv)~estimate the inverse-loss tail index against $\beta=1$;
(v)~locate the safe transverse geometry; and
(vi)~validate attainability with an explicit MLE.}

\emph{Average response and session loss (i)--(ii).}
The conventional descriptor is computed first: maximizing the latent average
QFI $\E_\xi[F_Q(T,\xi)]$ over interrogation time selects the standard Ramsey
working point, here the longer $T_{\rm avg}=2.11\,\mu$s. For each of the
$N_{\rm sess}=32000$ sessions a frozen configuration $\xi_i$ of
Eq.~\eqref{eq:nv-latent-config} is drawn (a sparse set of slow surface spins
whose quasi-static fields are the magnetic noise the diamond sees on that
run), and a calibration block at phases $\varphi=0,\pi$ estimates the session
contrast $\widehat C_i(T_k)$ and hence the conditional Fisher estimate
\begin{equation}
  \widehat F_i(T_k)=\kappa_{\rm ro}T_k^2\widehat C_i^2(T_k),
  \qquad
  \kappa_{\rm ro}=c_{\rm ro}^2,
  \label{eq:nv-calibrated-fisher}
\end{equation}
with $\kappa_{\rm ro}$ the readout Fisher efficiency, giving the session loss
sample $\widehat\ell_i=1/\widehat F_i$, or
$1/\widehat F_{\theta|\mu,i}$ when a nuisance is fitted explicitly (the
nuisance-projected Holevo loss $C^H_{B\theta|\mu}$ of Sec.~\ref{sec:transverse}
for a vector target). Sensing shots at $\varphi=\pi/2$ are estimated by MLE,
replaced by the profile estimator (covariance governed by $F_{\theta|\mu}^{-1}$)
when the nuisance is profiled out.

\emph{Tail certification, and the ranking reverses (iii).}
The session losses are tail-certified,
\begin{equation}
  \widehat{\mathcal{I}}^{\rm TC}_{\alpha}
  =
  \left[\widehat{\rm CVaR}_{\alpha}\!\left(\{\widehat\ell_i\}\right)\right]^{-1},
  \label{eq:nv-empirical-itc}
\end{equation}
with, for sorted losses,
\begin{equation}
  \widehat{\rm CVaR}_{\alpha}
  =
  \frac{1}{N_{\rm sess}-k_\alpha}
  \sum_{j=k_\alpha+1}^{N_{\rm sess}}\widehat\ell_{(j)},
  \qquad
  k_\alpha=\lfloor\alpha N_{\rm sess}\rfloor ,
  \label{eq:nv-empirical-cvar}
\end{equation}
and the ranking reverses. At $\alpha=0.95$ the empirical tail-certified
information is $4.15\times10^{-6}$ for the average-QFI design, against
$1.27\times10^{-2}$, $3.88\times10^{-3}$, and $6.27\times10^{-3}$ for the
safe-window, portfolio, and reserve designs, a recovery of roughly
$9.4\times10^{2}$ to $3.1\times10^{3}$ at the same shot budget and latent
ensemble. The same calibration data that make the average-QFI Ramsey design
look optimal expose it as operationally worst once the latent environment is
treated as quenched. Session-bootstrap $90\%$ intervals for
$\mathcal{I}^{\rm TC}_{0.95}$ are
$[1.89\times10^{-6},1.74\times10^{-5}]$,
$[1.24\times10^{-2},1.29\times10^{-2}]$,
$[3.56\times10^{-3},4.27\times10^{-3}]$, and
$[6.08\times10^{-3},6.47\times10^{-3}]$; the $\alpha=0.95$ CVaR averages the
$1600$ worst tail sessions.

\emph{Tail exponent and safe geometry (iv)--(v).}
A finite empirical CVaR is reported with the inverse-loss tail index that produced it. For a
threshold $u$ with $m=\#\{i:\widehat\ell_i>u\}$ exceedances, the Hill
diagnostic~\cite{Hill1975,Embrechts1997,deHaanFerreira2006}
\begin{equation}
  \widehat\beta(u)
  =
  \left[\frac1m\sum_{\widehat\ell_i>u}
  \log\!\left(\frac{\widehat\ell_i}{u}\right)\right]^{-1},
  \label{eq:hill-tail-estimator}
\end{equation}
with ideal standard error $\widehat\beta/\sqrt m$, places each design relative
to $\beta=1$ (Fig.~\ref{fig:experimental-certification}, right); the value is
intrinsic only when the lower one-sided bound clears $\beta=1$ or a physical
reserve gives $F_\perp(\xi)\geq F_0>0$ over the certified support, otherwise it
must be accompanied by the regulator flow of Sec.~\ref{sec:transition}. The
protocol also records where the safe geometry lies: the first latent
Fisher-zero boundary is $T_{\rm safe}=\pi/(2\Lambda)=1.57\,\mu$s, the
average-QFI time $T_{\rm avg}=2.11\,\mu$s sits \emph{beyond} it inside the
Fisher-glass region, and the tail-certified optimizer selects
$T_{\rm TC}=0.95\,\mu$s before it, with the safe phase, Fisher angle $\psi$, and
transverse gain $g_\perp$ of Secs.~\ref{sec:transverse}--\ref{sec:scaling}
separating a certifiable schedule from a catastrophic one.

\emph{Physical repair and attainability (vi).}
The safe-window, portfolio, and reserve schedules are exactly the repairs of
Sec.~\ref{sec:repair} instantiated on hardware; the remaining repairs and the
extrinsic regulators of Secs.~\ref{sec:transition} and~\ref{sec:repair} attach
at the same point. Estimator attainability is confirmed by repeating the sensing
block $R$ times at fixed calibrated curvature and forming
\begin{equation}
  \mathcal A_{\rm D}(i)
  =
  n\,\widehat{\rm Var}_{r}(\widehat\theta_{ir})\,
  \widehat F_{i,\theta|\mu},
  \label{eq:nv-attainability-ratio}
\end{equation}
where $r=1,\dots,R$ indexes repeated sensing blocks at fixed $\xi_i$,
$\widehat\theta_{ir}$ is the MLE in repeat $r$, $\widehat{\rm Var}_r$ is the
sample variance over those repeats, and $n$ is the sensing-shot budget. This
ratio concentrates near unity in the certified region while the average-QFI
design is exposed by its inverse-loss tail. Sensitivity sweeps, finite-session
convergence, Hill-exponent intervals, and a direct Fisher-loss versus MLE-error
comparison are reported in the Supplemental Methods
(Appendix~\ref{app:finite-sample},
Figs.~\ref{fig:nv-sensitivity-sweeps} and \ref{fig:nv-robustness-checks}).

\begin{table*}[tp]
\centering
\small
\renewcommand{\arraystretch}{1.25}
\newcommand{\scopecell}[2]{\parbox[t]{#1}{\raggedright\arraybackslash #2}}
\caption{Scope of the main statements.}
\label{tab:scope-assumptions}
\begin{tabular}{@{}lll@{}}
\hline\hline
\scopecell{0.24\textwidth}{Statement} &
\scopecell{0.46\textwidth}{Main assumptions} &
\scopecell{0.24\textwidth}{Status}\\
\hline
\scopecell{0.24\textwidth}{CVaR-QCRB and tail-certified loss}
&
\scopecell{0.46\textwidth}{Session-fixed latent configuration; local
regularity; pointwise QCRB or Holevo bound}
&
\scopecell{0.24\textwidth}{General certification bound}\\

\scopecell{0.24\textwidth}{Nuisance-projected scalar resource}
&
\scopecell{0.46\textwidth}{Regular nuisance block or Moore--Penrose support
convention}
&
\scopecell{0.24\textwidth}{Exact local scalar/directional statement}\\

\scopecell{0.24\textwidth}{Geometric Fisher-zero exponent}
&
\scopecell{0.46\textwidth}{Smooth latent density; codimension-$c$ zero; local
behavior $F\asymp r^{2m}$}
&
\scopecell{0.24\textwidth}{Geometric tail law}\\

\scopecell{0.24\textwidth}{Loschmidt-zero boundary}
&
\scopecell{0.46\textwidth}{Pure-dephasing conditional coherence with real
zeros in latent support}
&
\scopecell{0.24\textwidth}{Model-specific mechanism for Fisher zeros}\\

\scopecell{0.24\textwidth}{Portfolio exponent}
&
\scopecell{0.46\textwidth}{Positive-channel decomposition; zero-nondegenerate
active arms}
&
\scopecell{0.24\textwidth}{Sparse/quasi-static tail law}\\

\scopecell{0.24\textwidth}{Fisher reserve}
&
\scopecell{0.46\textwidth}{Positive safe reference channel over the certified
latent support}
&
\scopecell{0.24\textwidth}{Configuration-wise sufficient repair}\\

\scopecell{0.24\textwidth}{Action-separation scaling}
&
\scopecell{0.46\textwidth}{Fixed safe latent phase; smooth contrast factors
without adverse scaling}
&
\scopecell{0.24\textwidth}{Scaling criterion}\\

\scopecell{0.24\textwidth}{Shallow-NV simulation}
&
\scopecell{0.46\textwidth}{Sparse slow RTN surface fluctuators and calibrated
Ramsey contrast}
&
\scopecell{0.24\textwidth}{Representative platform demonstration}\\
\hline\hline
\end{tabular}
\end{table*}

% ============================================================
\section{Discussion and scope}
\label{sec:conclusion}
% ============================================================

We have shown that latent non-self-averaging sensors require certification of
the session-resolved attainable loss, not only the averaged Fisher response.
The obstruction is not low average sensitivity but rare latent configurations
in which usable information vanishes or becomes tangent to nuisance motion, and
in general it cannot be inferred from averaged Fisher geometry
(Theorem~\ref{thm:no-average}), and is moreover not even a renormalized
Fisher matrix (Sec.~\ref{sec:no-effective}). The certificate is governed by a
sharp integrability transition at $\beta=1$ (Theorem~\ref{thm:fisher-glass}),
the certified quantum resource is transverse to latent disorder
(Theorem~\ref{thm:transverse}), and the repairs are geometric: zero-free
windows, codimension raised by nondegenerate portfolios, configuration-wise
reserves, signal--disorder action separation, and the weakest target-identifying
Fisher cut (Theorem~\ref{thm:fisher-cut}). The shallow-NV tournament shows that
this certificate is experimentally estimable with finite calibration and
sensing shots, and that it reverses the protocol ranking selected by average
QFI.

Outside these assumptions the same six-step certification protocol of
Sec.~\ref{sec:nv-platform} remains diagnostic, with the explicit exponents and
zero boundaries then model-dependent rather than universal. The boundary of the
theory is also its reassurance: in self-averaging environments average Fisher
information may remain the right design objective, whereas in latent
non-self-averaging environments quantum enhancement must pass the stricter test
of the quenched pipeline of Eq.~\eqref{eq:intro-pipeline}. The certificate of
quantum metrological advantage is then a property of the quenched Fisher
geometry, not of any averaged channel.

\section{Author contributions}

E.M.M. conceived the project, developed the theoretical model, performed the
numerical simulations, and wrote the manuscript. K.A. reviewed the manuscript,
provided logistical support, and supervised the project. All authors contributed
to the interpretation of the results and the finalization of the manuscript.

\begin{acknowledgments}
This research was supported by JSPS KAKENHI Grant Number 24K21730.
\end{acknowledgments}

% ============================================================
\FloatBarrier
\appendix

\section*{Supplemental Methods overview}

The Supplemental Methods provide the proof details, model-specific derivations,
finite-regularization laws, additional platform robustness figures, and
simulation algorithm supporting the main-text statements. Appendix~%
\ref{app:comparison} places the framework among neighboring metrological risk
notions; Appendices~\ref{app:cvar}--\ref{app:noncommutation} prove the
certification bound and the no-averaged-Fisher theorem; Appendix~%
\ref{app:holevo} develops the Holevo target-map extension; Appendices~%
\ref{app:fisher-cut-proof}--\ref{app:action-separation-proof} prove the
architecture, codimension, Loschmidt-zero, and action-separation laws;
Appendices~\ref{app:rtn}--\ref{app:dc-bound} give the platform-specific
derivations; and Appendices~\ref{app:finite-sample}--\ref{app:regulator-flow}
document the simulation and the regulator flow.

\section{Comparison with neighboring metrological frameworks}
\label{app:comparison}

The distinction from neighboring frameworks is the order forced by frozen
latent sessions: condition on the session, project nuisance directions, form
the attainable inverse loss, and only then certify its latent-session tail.
Table~\ref{tab:risk-comparison} summarizes the comparison.

\begin{table*}[t]
\centering
\caption{Comparison with neighboring metrological risk notions.}
\label{tab:risk-comparison}
\newcommand{\riskcmpcell}[2]{\parbox[t]{#1}{\raggedright #2}}
\begin{tabular}{@{}llll@{}}
\hline
\riskcmpcell{0.20\textwidth}{Framework} &
\riskcmpcell{0.23\textwidth}{Random object} &
\riskcmpcell{0.25\textwidth}{Risk/loss} &
\riskcmpcell{0.24\textwidth}{Fisher-glass distinction}\\
\hline
\riskcmpcell{0.20\textwidth}{Average-QFI or decoherence-limited metrology}
&
\riskcmpcell{0.23\textwidth}{Self-averaged or fixed noisy channel}
&
\riskcmpcell{0.25\textwidth}{Average QFI / classical Fisher information (FI)}
&
\riskcmpcell{0.24\textwidth}{Can miss rare sessions where projected Fisher
information nearly vanishes.}\\

\riskcmpcell{0.20\textwidth}{Nuisance-parameter metrology}
&
\riskcmpcell{0.23\textwidth}{Fixed local model with nuisance coordinates}
&
\riskcmpcell{0.25\textwidth}{Schur-complement or Holevo nuisance loss}
&
\riskcmpcell{0.24\textwidth}{Adds a quenched distribution of such losses
across sessions.}\\

\riskcmpcell{0.20\textwidth}{Bayesian metrology}
&
\riskcmpcell{0.23\textwidth}{Prior over target and/or nuisance variables}
&
\riskcmpcell{0.25\textwidth}{Prior-averaged loss}
&
\riskcmpcell{0.24\textwidth}{Here $\theta$ is fixed locally; the risk is over
frozen session labels.}\\

\riskcmpcell{0.20\textwidth}{Minimax robust metrology}
&
\riskcmpcell{0.23\textwidth}{Worst-case model element}
&
\riskcmpcell{0.25\textwidth}{Supremum loss}
&
\riskcmpcell{0.24\textwidth}{CVaR weights bad latent sessions by probability
rather than only by worst case.}\\

\riskcmpcell{0.20\textwidth}{Compound/random-channel metrology}
&
\riskcmpcell{0.23\textwidth}{Family or mixture of channels}
&
\riskcmpcell{0.25\textwidth}{Channel-family or averaged-channel criterion}
&
\riskcmpcell{0.24\textwidth}{The loss is formed after conditioning on the
frozen session, not on the annealed channel.}\\

\riskcmpcell{0.20\textwidth}{Finite-sample or outage-style metrology}
&
\riskcmpcell{0.23\textwidth}{Estimator outcome at finite sample size}
&
\riskcmpcell{0.25\textwidth}{Success probability, outage probability, or
finite-sample accuracy}
&
\riskcmpcell{0.24\textwidth}{Tests finite-shot attainability of a target error;
Fisher-glass risk is over session-to-session latent geometry and is applied to
the conditional loss before estimator-error validation.}\\

\riskcmpcell{0.20\textwidth}{Fisher-glass certification}
&
\riskcmpcell{0.23\textwidth}{Frozen latent configuration $\xi$}
&
\riskcmpcell{0.25\textwidth}{$[\CVaR_\alpha(\ell_{\mathcal P}(\xi))]^{-1}$}
&
\riskcmpcell{0.24\textwidth}{Condition, project nuisance, invert, then
tail-certify.}\\
\hline
\end{tabular}
\end{table*}

\section{CVaR preliminaries and proof of the CVaR quantum Cram\'er--Rao bound}
\label{app:cvar}

For a conditional local model with QFI matrix $J(\xi)$, the directional
inverse-Fisher loss for a target direction $w$ is
\begin{equation}
  \ell_w(\xi)
  =
  \begin{cases}
  w^\top J(\xi)^+w, & w\in\operatorname{range}J(\xi),\\
  +\infty, & w\notin\operatorname{range}J(\xi).
  \end{cases}
  \label{eq:extended-loss}
\end{equation}
Equivalently, the corresponding directional Fisher information is
\begin{equation}
  F_w(\xi)=\ell_w(\xi)^{-1}
  =
  \min_{\phi:\,w^\top\phi=1}\phi^\top J(\xi)\phi ,
  \label{eq:directional-variational}
\end{equation}
with $F_w=0$ when the direction is locally unidentifiable. This is the
formal convention behind the scalar and directional tail-certified
information used in the main text.

CVaR has the dual representation
\begin{equation}
  \begin{aligned}
  \CVaR_\alpha(X)
  &=
  \sup_{h\in\mathcal Q_\alpha}\E[hX],\\
  \mathcal Q_\alpha
  &=
  \{h:0\leq h\leq(1-\alpha)^{-1},\ \E h=1\}.
  \end{aligned}
  \label{eq:cvar-dual}
\end{equation}
Monotonicity follows immediately. The pointwise quantum Cram\'er--Rao bound
gives $R(\xi;\theta)\geq1/[nF_Q(\theta,\xi)]$ for every latent
configuration where local unbiasedness and regularity hold. Applying CVaR
monotonicity gives Eq.~\eqref{eq:cvar-qcrb}. The multiparameter result follows
from the pointwise bound
$w^\top V(\xi;\theta)w\geq n^{-1}\ell_w(\xi)$, with the extended convention
of~\eqref{eq:extended-loss}. If $w$ is outside the range of $J(\xi)$, the
target direction is locally unidentifiable for that configuration and the
right-hand side is $+\infty$.

For the non-matricial tail-certified body statement, duality gives
\begin{equation}
  \CVaR_\alpha(w^\top J(\xi)^{-1}w)
  =
  \sup_{h\in\mathcal Q_\alpha}
  w^\top\E[h(\xi)J(\xi)^{-1}]w .
  \label{eq:body-support}
\end{equation}
This is the supremum of positive quadratic forms, hence convex and
homogeneous of degree two. Its unit sublevel set is symmetric and convex.
It is not generally quadratic: for two equally likely configurations with
$J_1^{-1}=\operatorname{diag}(1,100)$ and
$J_2^{-1}=\operatorname{diag}(100,1)$ at $\alpha=1/2$, the risk function is
$R(w)=\max\{w^\top J_1^{-1}w,w^\top J_2^{-1}w\}$. It violates the
parallelogram identity and therefore cannot arise from a single matrix.

\section{Noncommutation of latent certification}
\label{app:noncommutation}

The distinction from ordinary nuisance projection and ordinary risk
optimization can be stated as a noncommutation property. Let $J(\xi)$ be a
latent session-resolved Fisher matrix and define the projected scalar
information
\begin{equation}
  F_{\theta|\mu}(\xi)
  =
  J_{\theta\theta}(\xi)
  -
  J_{\theta\mu}(\xi)J_{\mu\mu}(\xi)^{-1}J_{\mu\theta}(\xi).
  \label{eq:methods-noncommutation-projected-fisher}
\end{equation}
Then the tail-certified information
\begin{equation}
  \mathcal{I}^{\rm TC}_{\alpha,\theta|\mu}
  =
  \left[
  \CVaR_\alpha\!\left(F_{\theta|\mu}(\xi)^{-1}\right)
  \right]^{-1}
  \label{eq:methods-noncommutation-itc}
\end{equation}
is not determined by the averaged Fisher matrix
$\overline J=\E_\xi[J(\xi)]$. In particular, there exist latent ensembles
with the same $\overline J$ and the same average projected information
$\E_\xi[F_{\theta|\mu}(\xi)]$ but different tail-certified information,
including one with $\mathcal{I}^{\rm TC}_{\alpha,\theta|\mu}=0$ and one with
$\mathcal{I}^{\rm TC}_{\alpha,\theta|\mu}>0$.

It is enough to consider the scalar case and then embed the construction as
the $\theta$ block of a larger Fisher matrix. One ensemble has
$F(\xi)=1$ deterministically, so its tail-certified information is finite. A
second ensemble can be chosen with bounded support, with density nonzero near
$F=0$ so that $F^{-1}$ has a nonintegrable upper tail, and with compensating
mass at larger finite $F$ so that $\E F=1$ remains finite. The averaged
Fisher matrix and average projected information are the same, but the CVaR
of $F^{-1}$ is infinite in the second ensemble. Thus latent certification
depends on the projected inverse-loss tail, not on the averaged geometry.

Concretely, take ensemble~A with $F_A=1$ deterministically, so $\E F_A=1$ and
$\E[1/F_A]=1$ (finite certificate). For ensemble~B, let $F_B$ be, with
probability $\tfrac12$, a draw from the density $p(F)=\tfrac12 F^{-1/2}$ on
$(0,1]$, and with probability $\tfrac12$ a point mass at $F_0=5/3$. Then
\begin{equation}
  \E F_B=\tfrac12\!\int_0^1\! F\,p(F)\,dF+\tfrac12 F_0
  =\tfrac12\cdot\tfrac13+\tfrac12\cdot\tfrac53=1,
\end{equation}
matching ensemble~A, whereas
\begin{equation}
  \E[1/F_B]\ \ge\ \tfrac12\!\int_0^1\! F^{-1}p(F)\,dF
  =\tfrac14\!\int_0^1\! F^{-3/2}\,dF=\infty .
\end{equation}
Hence $\CVaR_\alpha(1/F_B)=\infty$ and the certificate is zero, while ensembles
A and B share $\E F=1$ (the same averaged Fisher matrix and averaged projected
information) yet differ in tail-certified information, one finite and one zero.

\section{Holevo tail-certified extension}
\label{app:holevo}

The symmetric logarithmic derivative (SLD) Fisher matrix need not give an
attainable multiparameter bound when different target components require
incompatible measurements. The attainable local asymptotic replacement is the
Holevo loss after nuisance projection.

Let
\begin{equation}
  \omega=(\theta,\mu),\qquad
  \theta\in\mathbb R^p,\qquad
  \mu\in\mathbb R^q,
  \label{eq:holevo-local-parameters}
\end{equation}
and let $\rho_\omega^\xi$ be the conditional local model selected by the latent
configuration $\xi$. The reported target need not be the full vector
$\theta$. We write
\begin{equation}
  \vartheta=B\theta,\qquad
  B\in\mathbb R^{d_B\times p},
  \label{eq:methods-reported-target}
\end{equation}
where $B$ has full row rank, and use a positive definite weight
$W_B\succ0$ on the reported target. Full-vector estimation corresponds to
$B=I_p$; directional estimation corresponds to $d_B=1$ and $B=w^\top$.
Equivalently, a rank-deficient weight $W\succeq0$ on the original
$\theta$-space is represented by choosing $B$ and $W_B$ such that
$W=B^\top W_BB$. The estimator is then required to be locally unbiased only
for the reported target $B\theta$.

For fixed $\xi$, write $\rho=\rho_{\omega_0}^\xi$ and
$\dot\rho_\eta=\partial_\eta\rho_\omega^\xi|_{\omega_0}$. Define the admissible
observable class
\begin{equation}
\begin{aligned}
\mathcal X_{B|\mu}(\xi)
=
\bigl\{
&X=(X_a)_{a=1}^{d_B}: X_a=X_a^\dagger,\\
&\Tr(\rho X_a)=0,\\
&\Tr(\dot\rho_{\theta_j}X_a)=B_{aj},\\
&\Tr(\dot\rho_{\mu_\ell}X_a)=0
\bigr\}.
\end{aligned}
  \label{eq:holevo-admissible-class}
\end{equation}
For $X\in\mathcal X_{B|\mu}(\xi)$, set
\begin{equation}
  Z_{ab}(X;\xi)=\Tr\!\left(\rho X_aX_b\right).
  \label{eq:holevo-z-matrix}
\end{equation}
The nuisance-projected Holevo loss for the reported target is
\begin{equation}
\begin{aligned}
C^H_{B\theta|\mu}(W_B;\xi)
=
\inf_{X\in\mathcal X_{B|\mu}(\xi)}
\biggl\{
&\operatorname{tr}\!\left[W_B\operatorname{Re}Z_X\right]\\
&+
\left\|
W_B^{1/2}\operatorname{Im}Z_XW_B^{1/2}
\right\|_1
\biggr\}.
\end{aligned}
  \label{eq:methods-holevo-loss}
\end{equation}
Here $Z_X$ abbreviates $Z(X;\xi)$ inside the infimum.
If $\mathcal X_{B|\mu}(\xi)=\varnothing$, or if a $W_B$-visible reported
target direction is locally indistinguishable from nuisance motion, we set
\begin{equation}
  C^H_{B\theta|\mu}(W_B;\xi)=+\infty.
  \label{eq:methods-holevo-infinite}
\end{equation}
The constraint $\Tr(\dot\rho_{\mu_\ell}X_a)=0$ is the nuisance-projection
condition: the reported estimator is locally insensitive to nuisance
displacements, while the nuisance remains part of the local statistical
experiment.

Under the usual regularity assumptions for the Holevo bound, and when the
completed local model is nonsingular, Eq.~\eqref{eq:methods-holevo-loss}
agrees with the corresponding zero-weight construction. In particular, for
$B=I_p$ and $W\succ0$,
\begin{equation}
C^H_{\theta|\mu}(W;\xi)
=
\lim_{\varepsilon\downarrow0}
C^H_{\theta,\mu}
\!\left(
\begin{pmatrix}
W&0\\
0&\varepsilon I_q
\end{pmatrix}
;\xi
\right).
  \label{eq:methods-holevo-zero-weight}
\end{equation}
Here $C^H_{\theta,\mu}$ denotes the ordinary Holevo loss for the completed
target $(\theta,\mu)$. For a general target map $B$, one may complete $B$ to
a locally invertible
change of target coordinates and assign vanishing weight to the unreported
coordinates and to $\mu$. The constrained definition
\eqref{eq:methods-holevo-loss} is the primary definition and remains
meaningful when unreported target directions are not locally identifiable.

For any locally asymptotically covariant estimator sequence of the reported
target, with conditional covariance $V_{B,n}(\xi)$, assume the Holevo
remainder is uniform over the safe latent support, or more generally that its
negative part is $o(n^{-1})$ in CVaR. Then
\begin{equation}
\begin{aligned}
&\liminf_{n\to\infty}
  n\,\CVaR_\alpha\!\left(
  \Tr\,W_B V_{B,n}(\xi)
  \right)\\
&\qquad\geq
  \CVaR_\alpha\!\left(
  C^H_{B\theta|\mu}(W_B;\xi)
  \right).
\end{aligned}
  \label{eq:methods-holevo-bound}
\end{equation}
If the conditional model is quantum local asymptotic normality (QLAN)-regular
and the Holevo bound is uniformly
attainable over the safe latent support, with a CVaR-controlled remainder,
then
\begin{equation}
\begin{aligned}
&\lim_{n\to\infty}
  n\,\CVaR_\alpha\!\left(
  \Tr\,W_B V_{B,n}(\xi)
  \right)\\
&\qquad=
  \CVaR_\alpha\!\left(
  C^H_{B\theta|\mu}(W_B;\xi)
  \right).
\end{aligned}
  \label{eq:methods-holevo-cvar}
\end{equation}
Under these conditions the bound is tight at the level of the leading
$n^{-1}$ loss~\cite{YangChiribellaHayashi2019}. The vector-target
tail-certified Holevo information is therefore
\begin{equation}
\mathcal{I}^{\rm TC,H}_{\alpha,B|\mu}(W_B)
=
\left[
\CVaR_\alpha
\left(
C^H_{B\theta|\mu}(W_B;\xi)
\right)
\right]^{-1}.
  \label{eq:methods-holevo-closure}
\end{equation}
Thus the vector-target Fisher-glass criterion is the tail integrability of
the conditional attainable loss $C^H_{B\theta|\mu}(W_B;\xi)$, not the average
SLD QFI matrix.

Let the SLD Fisher matrix of the full local model be block-decomposed as
\begin{equation}
J(\xi)=
\begin{pmatrix}
J_{\theta\theta} & J_{\theta\mu}\\
J_{\mu\theta} & J_{\mu\mu}
\end{pmatrix}.
  \label{eq:methods-holevo-block-j}
\end{equation}
The nuisance-projected SLD Fisher matrix is
\begin{equation}
J_{\theta|\mu}
=
J_{\theta\theta}
-
J_{\theta\mu}J_{\mu\mu}^{+}J_{\mu\theta},
  \label{eq:methods-holevo-schur}
\end{equation}
with the usual Moore--Penrose support convention on singular strata. The
compatible SLD loss for the reported target is
\begin{equation}
C^S_{B\theta|\mu}(W_B;\xi)
=
\begin{cases}
\operatorname{tr}\!\left[
W_B\,B J_{\theta|\mu}^{+}B^\top
\right],
&
\begin{gathered}
\operatorname{row}(B)\subseteq\\
\operatorname{range}(J_{\theta|\mu}),
\end{gathered}
\\[1ex]
+\infty,
&
\text{otherwise.}
\end{cases}
  \label{eq:methods-holevo-compatible-loss}
\end{equation}
The Holevo loss satisfies
\begin{equation}
  C^H_{B\theta|\mu}(W_B;\xi)\ge C^S_{B\theta|\mu}(W_B;\xi).
  \label{eq:methods-holevo-sld-lower}
\end{equation}
If the nuisance-projected reported-target model is SLD-compatible, the
inequality is saturated:
\begin{equation}
C^H_{B\theta|\mu}(W_B;\xi)
=
\operatorname{tr}\!\left[
W_B\,B J_{\theta|\mu}^{+}B^\top
\right].
  \label{eq:methods-holevo-compatible}
\end{equation}
For a scalar reported target, $d_B=1$, $B=w^\top$, $W_B=1$, this gives
\begin{equation}
  C^H_{w^\top\theta|\mu}(1;\xi)=w^\top J_{\theta|\mu}^{+}w,
  \label{eq:methods-holevo-directional}
\end{equation}
and for the original one-dimensional scalar case,
\begin{equation}
  C^H_{\theta|\mu}(1;\xi)=F_{\theta|\mu}^{-1}(\xi).
  \label{eq:methods-holevo-scalar}
\end{equation}

When the compatible loss is finite and nonzero, define
\begin{equation}
\kappa_H(B,W_B;\xi)
=
\frac{
C^H_{B\theta|\mu}(W_B;\xi)
}{
\operatorname{tr}\!\left[
W_B\,B J_{\theta|\mu}^{+}B^\top
\right]
}
\ge1.
  \label{eq:methods-holevo-kappa}
\end{equation}
This separates projected identifiability loss from genuine measurement
incompatibility. If near a Holevo-null set
\begin{equation}
  C^H_{B\theta|\mu}(W_B;s,u)\asymp r^{-\sigma}
  \label{eq:methods-holevo-local-tail}
\end{equation}
in codimension $c$ (with $\sigma=2m$ for a Fisher zero of order $m$, matching
the $2m$ of Eq.~\eqref{eq:codim-exponent}), and the latent density scales as $r^\nu$, then
\begin{equation}
\Pr\!\left(C^H_{B\theta|\mu}(W_B;\xi)>x\right)
\asymp
x^{-(c+\nu)/\sigma}.
  \label{eq:methods-holevo-tail-law}
\end{equation}
Therefore finite vector-target tail certification requires
\begin{equation}
  \frac{c+\nu}{\sigma}>1.
  \label{eq:methods-holevo-integrability}
\end{equation}
For a simple codimension-one Fisher zero with bounded $\kappa_H$, $\sigma=2$,
hence the inverse-loss survival exponent is $1/2$, and all finite-level CVaR
losses diverge.

As a minimal vector example, consider the local pure-qubit model
\begin{equation}
|\psi_{\theta,\xi}\rangle
=
|0\rangle
+
\frac{a(\xi)}{2}(\theta_1+i\theta_2)|1\rangle
+
O(|\theta|^2).
  \label{eq:methods-holevo-qubit}
\end{equation}
At $\theta=0$, $J_{\theta\theta}=a(\xi)^2I_2$. For the full two-component
target, $B=I_2$, $W_B=I_2$, the compatible SLD loss is $2/a(\xi)^2$. The
two-parameter pure-qubit Holevo--Nagaoka bound gives
\begin{equation}
  C^H_{\theta}(I_2;\xi)=\frac{4}{a(\xi)^2},
  \label{eq:methods-holevo-qubit-full}
\end{equation}
so $\kappa_H=2$. For a one-dimensional reported direction $B=w^\top$, the
problem is scalar and
\begin{equation}
  C^H_{w^\top\theta}(1;\xi)=\frac{|w|^2}{a(\xi)^2}.
  \label{eq:methods-holevo-qubit-direction}
\end{equation}
If an unknown nuisance coordinate shifts only the first quadrature,
\begin{equation}
|\psi_{\theta,\mu,\xi}\rangle
=
|0\rangle
+
\frac{a(\xi)}{2}\bigl((\theta_1+\mu)+i\theta_2\bigr)|1\rangle
+
O(|\theta|^2+|\mu|^2),
  \label{eq:methods-holevo-qubit-nuisance}
\end{equation}
then $\dot\rho_{\theta_1}^\xi=\dot\rho_{\mu}^\xi$. The full target
$(\theta_1,\theta_2)$ is not identifiable after nuisance projection, so
$C^H_{\theta|\mu}(I_2;\xi)=+\infty$. However the reported target
$B=e_2^\top$ remains identifiable, with admissible observable
$X=\sigma_y/a(\xi)$ and
\begin{equation}
  C^H_{e_2^\top\theta|\mu}(1;\xi)=\frac{1}{a(\xi)^2}.
  \label{eq:methods-holevo-qubit-nuisance-direction}
\end{equation}
Thus the target-map formulation distinguishes an unidentifiable full target
from an identifiable reported target.

Quantum nuisance-parameter treatments and practical evaluations of the
Holevo bound provide the corresponding tools for vector targets
~\cite{SuzukiYangHayashi2020,AlbarelliFrielRobertsonDatta2019}.

\section{Proof of the Fisher-cut theorem}
\label{app:fisher-cut-proof}

For a cut $S$, there exists $\phi_S$ with $a_e^\top\phi_S=0$ for
$e\notin S$ and $w^\top\phi_S\neq0$. Normalize $w^\top\phi_S=1$. Then
$F_w\leq\sum_{e\in S}g_e(a_e^\top\phi_S)^2\leq
C_S\sum_{e\in S}g_e$, and minimizing over cuts gives the upper bound.

For the lower bound, consider every subset $T\subseteq E$ spanning $w$ and
choose coefficients $c_e^{(T)}$ with $w=\sum_{e\in T}c_e^{(T)}a_e$. Since
there are finitely many spanning subsets, choose $\delta>0$ so that
$\sum_{e\in T}|c_e^{(T)}|\delta<1$ for all of them. For any feasible $\phi$,
let $S_\phi=\{e:|a_e^\top\phi|\geq\delta\}$. If $E\setminus S_\phi$ spanned
$w$, then $|w^\top\phi|<1$, contradiction. Thus $S_\phi$ is a cut, and
$\sum_e g_e(a_e^\top\phi)^2\geq\delta^2\Phi_w$. Minimizing over $\phi$ gives
the lower bound.

For a fixed cut $S$, independence and
$\Prob(g_e<\varepsilon)=\varepsilon^{\eta_e+o(1)}$ imply
$\Prob(\sum_{e\in S}g_e<\varepsilon)=
\varepsilon^{\sum_{e\in S}\eta_e+o(1)}$. A finite union over cuts is
dominated by the smallest exponent. Setting $\varepsilon=x^{-1}$ proves the
tail law.

\section{Nuisance-projected Fisher cuts and transverse Fisher volume}
\label{app:transverse-volume}

This appendix records the multiparameter geometry behind the
signal--disorder action-separation law. It should be read as a regular-stratum
statement: the nuisance block is assumed nonsingular on the configurations
under consideration, while the extended inverse-Fisher convention applies
when the target direction is not identifiable.

\begin{proposition}[Nuisance-projected Fisher cuts]
\label{prop:transverse-fisher-cuts}
Consider one target coordinate $\theta$ and $q$ nuisance coordinates $\mu$.
Let
\begin{equation}
  \begin{aligned}
  J(\xi)&=\sum_{e\in E}g_e(\xi)b_eb_e^\top,\qquad g_e(\xi)\geq0,\\
  b_e&=(s_e,n_e^\top)\in\mathbb R^{q+1}.
  \end{aligned}
  \label{eq:transverse-channel}
\end{equation}
Assume the nuisance block is nonsingular on the relevant support. The
nuisance-projected Fisher information for $\theta$ is
\begin{equation}
  F_{\theta|\mu}(\xi)
  =
  J_{\theta\theta}
  -
  J_{\theta\mu}J_{\mu\mu}^{-1}J_{\mu\theta}.
  \label{eq:nuisance-schur}
\end{equation}
Let $B_S$ denote the $(q+1)\times(q+1)$ matrix with rows $b_e^\top$,
$e\in S$, and let $N_R$ denote the $q\times q$ matrix with rows $n_e^\top$,
$e\in R$. Then
\begin{equation}
  F_{\theta|\mu}
  =
  \frac{
  \sum_{|S|=q+1}
  \left(\prod_{e\in S}g_e\right)\det(B_S)^2
  }{
  \sum_{|R|=q}
  \left(\prod_{e\in R}g_e\right)\det(N_R)^2
  },
  \label{eq:nuisance-volume-ratio}
\end{equation}
with the empty determinant convention for $q=0$.

A nuisance-projected Fisher cut is a subset $C\subset E$ such that, after
removing $C$, the target response of every remaining channel can be absorbed
into the nuisance span:
\begin{equation}
  \begin{aligned}
  C\in\mathcal C_{\theta|\mu}
  \Longleftrightarrow\;&
  \exists u_C\in\mathbb R^q\ \text{such that}\\
  &s_e+n_e^\top u_C=0
  \quad \text{for all }e\notin C .
  \end{aligned}
  \label{eq:nuisance-absorption-cut}
\end{equation}
Equivalently, the restricted signal vector lies in the range of the
restricted nuisance-response matrix.
Define
\begin{equation}
  \Phi_{\theta|\mu}(\xi)
  =
  \min_{C\in\mathcal C_{\theta|\mu}}\sum_{e\in C}g_e(\xi).
  \label{eq:nuisance-capacity}
\end{equation}
For fixed finite response vectors, on the same regular stratum, if the target
is identifiable before any channel failure, there are constants
$0<c_{\theta|\mu}<C_{\theta|\mu}<\infty$ such that
\begin{equation}
  c_{\theta|\mu}\Phi_{\theta|\mu}(\xi)
  \leq
  F_{\theta|\mu}(\xi)
  \leq
  C_{\theta|\mu}\Phi_{\theta|\mu}(\xi).
  \label{eq:nuisance-comparison}
\end{equation}
If the channel weights are independent and
\begin{equation}
  \Prob(g_e<\varepsilon)=\varepsilon^{\eta_e+o(1)},
  \qquad \varepsilon\to0,
  \label{eq:transverse-small-channel}
\end{equation}
then
\begin{equation}
  \Prob(F_{\theta|\mu}^{-1}>x)
  =
  x^{-\beta_{\theta|\mu}+o(1)},
  \qquad
  \beta_{\theta|\mu}
  =
  \min_{C\in\mathcal C_{\theta|\mu}}\sum_{e\in C}\eta_e .
  \label{eq:nuisance-tail}
\end{equation}
Thus finite nuisance-projected tail-certified information requires
$\beta_{\theta|\mu}>1$.
\end{proposition}

\paragraph*{Proof.}
The Schur complement identity
$F_{\theta|\mu}=\det J/\det J_{\mu\mu}$, together with Cauchy--Binet applied
to $J=D^\top GD$ and $J_{\mu\mu}=N^\top GN$, gives
\eqref{eq:nuisance-volume-ratio}. Equivalently,
\begin{equation}
  F_{\theta|\mu}
  =
  \min_{u\in\mathbb R^q}
  \sum_{e\in E}g_e(s_e+n_e^\top u)^2 .
  \label{eq:transverse-variational}
\end{equation}
For the upper bound, take a nuisance-projected Fisher cut $C$. By
definition, there exists $u_C$ such that $s_e+n_e^\top u_C=0$ for
$e\notin C$. Hence
\begin{equation}
  F_{\theta|\mu}
  \leq
  \sum_{e\in C}g_e(s_e+n_e^\top u_C)^2
  \leq
  C_C\sum_{e\in C}g_e .
\end{equation}
Minimizing over $C$ gives the upper bound in
\eqref{eq:nuisance-comparison}.

For the lower bound, finiteness of $E$ and fixed response vectors imply that
there is a $\delta>0$ such that whenever a subset $T\subset E$ is not
target-absorbable,
\begin{equation}
  \inf_u\max_{e\in T}|s_e+n_e^\top u|\geq\delta .
\end{equation}
For any $u$, let
$S(u)=\{e:|s_e+n_e^\top u|\geq\delta\}$. Then $E\setminus S(u)$ cannot
be nonabsorbable, since all its residuals are smaller than $\delta$.
Therefore $S(u)\in\mathcal C_{\theta|\mu}$. Thus
\begin{equation}
  \sum_e g_e(s_e+n_e^\top u)^2
  \geq
  \delta^2\sum_{e\in S(u)}g_e
  \geq
  \delta^2\Phi_{\theta|\mu} .
\end{equation}
Minimizing over $u$ gives the lower bound.

For a fixed cut $C$, independence and
\eqref{eq:transverse-small-channel} give
$\Prob(\sum_{e\in C}g_e<\varepsilon)
=\varepsilon^{\sum_{e\in C}\eta_e+o(1)}$. A finite union over
nuisance-projected cuts is dominated by the smallest exponent, and
\eqref{eq:nuisance-comparison} transfers the exponent to
$F_{\theta|\mu}$. Setting $\varepsilon=x^{-1}$ proves
\eqref{eq:nuisance-tail}.

\begin{corollary}[Critical transverse scaling]
\label{cor:critical-transverse-scaling}
Consider a family of sensors indexed by size $L$. Suppose that on a regular
nuisance stratum the projected signal gain and disorder action gain obey
\begin{equation}
  g_\perp(L)\sim L^{y_\perp},
  \qquad
  g_d(L)\sim L^{y_d}.
  \label{eq:app-critical-gains}
\end{equation}
Assume that the safe latent phase optimum remains $L$-independent, that the
smooth contrast factor has no adverse scaling, and that no additional
Fisher-zero singularity is introduced by the transverse angle. Then the
optimized per-shot tail-certified information scales as
\begin{equation}
  \mathcal{I}^{\rm TC}_{\alpha,\mathrm{shot}}(L)
  \sim
  L^{2(y_\perp-y_d)}.
  \label{eq:app-critical-shot}
\end{equation}
If $g_s(L)\sim L^{y_s}$ and $\sin^2\psi_L\sim L^{-\delta}$, then
$y_\perp=y_s-\delta/2$, so
\begin{equation}
  \mathcal{I}^{\rm TC}_{\alpha,\mathrm{shot}}(L)
  \sim
  L^{2(y_s-y_d)-\delta}.
  \label{eq:app-critical-angle}
\end{equation}
With negligible dead time, the information rate scales as
\begin{equation}
  \Gamma^{\rm TC}_{\alpha}(L)
  \sim
  L^{2y_\perp-y_d}.
  \label{eq:app-critical-rate}
\end{equation}
\end{corollary}

\paragraph*{Proof.}
At fixed safe latent phase $\Phi=g_d(L)T\Lambda$, the interrogation time
scales as $T=\Phi/[g_d(L)\Lambda]$. The projected Fisher scale is
$g_\perp^2(L)T^2$ times an $L$-independent dimensionless tail factor.
Therefore the inverse-CVaR scale is proportional to $(g_\perp/g_d)^2$. The
rate statement follows by dividing by $T$ in the negligible-dead-time limit.

\paragraph*{Generic volume-hypergraph consequence.}
When the nuisance response rows are in generic position, so that every subset
of at most $q$ nuisance rows has full row rank and every larger subset has
rank $q$, the target-absorption cuts are exactly the vertex covers of the
nonzero $(q+1)$-volume hypergraph. For $K$ generic channels in
$\mathbb R^{q+1}$, every $(q+1)$-tuple has nonzero Fisher volume. Destroying
all transverse volumes then requires leaving at most $q$ channels, so the
smallest cut has size $K-q$. For identical simple-zero channels,
$\eta_e=1/2$, and therefore
\begin{equation}
  \beta_{\theta|\mu}=\frac{K-q}{2}.
  \label{eq:generic-transverse-exponent}
\end{equation}
Finite CVaR requires $\beta_{\theta|\mu}>1$, hence
\begin{equation}
  K\geq q+3.
  \label{eq:generic-transverse-threshold}
\end{equation}
The case $q=0$ recovers the scalar portfolio threshold $K\geq3$; one
nuisance direction raises the generic threshold to $K\geq4$.

\section{Proof of the Fisher-codimension law}
\label{app:codim-proof}

The event $\ell>x$ is $F<x^{-1}$. Near $Z$, this is equivalent to
$r\lesssim A(s,\hat u)^{-1/(2m)}x^{-1/(2m)}$. The tube probability is
asymptotic to
\begin{equation}
  \begin{aligned}
  &\int_Z\int_{S^{c-1}}
  \int_0^{A(s,\hat u)^{-1/(2m)}x^{-1/(2m)}}\\
  &\hspace{1.2cm}
  b_0(s,\hat u)r^{c-1+\nu}\,dr\,d\hat u\,d\sigma(s),
  \end{aligned}
\end{equation}
which equals $A_\eta x^{-(c+\nu)/(2m)}$ with finite positive $A_\eta$. The moment
criterion follows from
$\E[\ell^s]=s\int_0^\infty x^{s-1}\Prob(\ell>x)\,dx$.

\section{Proof of the Loschmidt-zero transition}
\label{app:loschmidt-proof}

Let $u=A\lambda$ and suppose $\chi(u)=a(u-u_1)^m+o(|u-u_1|^m)$ at the first
positive real zero $u_1$. If $A\Lambda\geq u_1$, then
$\lambda_*=u_1/A$ lies in the latent support. Near $\lambda_*$,
\begin{equation}
  \begin{aligned}
  F_A(\lambda)
  &=
  A^2C_A^2|\chi(A\lambda)|^2\\
  &=
  C_A^2|a|^2|A|^{2m+2}
  |\lambda-\lambda_*|^{2m}
  +o(|\lambda-\lambda_*|^{2m}).
  \end{aligned}
\end{equation}
Therefore $1/F_A>x$ is equivalent, to leading order, to
$|\lambda-\lambda_*|<c_0x^{-1/(2m)}$ for a positive constant $c_0$. If the
coupling density is continuous and nonzero at $\lambda_*$, the probability
of this interval is asymptotic to $2p(\lambda_*)c_0x^{-1/(2m)}$ for an
interior zero and half this value at a support endpoint. Since
$1/(2m)\leq1/2<1$, the first inverse-Fisher moment and every CVaR level
$\alpha<1$ diverge. If $A\Lambda<u_1$, compactness of $[0,\Lambda]$ and
continuity of $\chi$ imply $\min_{\lambda\in[0,\Lambda]}|\chi(A\lambda)|>0$,
so $1/F_A$ is bounded.

\section{Proof of the signal--disorder action separation law}
\label{app:action-separation-proof}

For $\Phi=g_d(M)T\Lambda$ and $U=\lambda/\Lambda$,
the pure-dephasing Fisher information
$F_M(T,\lambda)=g_s(M)^2T^2C_M^2(T)|\chi(g_d(M)T\lambda)|^2$
can be rewritten as
\begin{equation}
  F_M(T,\lambda)
  =
  \left(\frac{g_s(M)}{g_d(M)}\right)^2
  \frac{C_M^2(T)}{\Lambda^2}
  \Phi^2|\chi(\Phi U)|^2 .
\end{equation}
The prefactor is deterministic with respect to the latent variable, so
positive homogeneity of CVaR gives
\begin{equation}
  \begin{aligned}
  \CVaR_\alpha\!\left(\frac{1}{F_M}\right)
  &=
  \left(\frac{g_d(M)}{g_s(M)}\right)^2
  \frac{\Lambda^2}{C_M^2(T)}\\
  &\quad\times
  \CVaR_\alpha\!\left(
  \frac{1}{\Phi^2|\chi(\Phi U)|^2}
  \right).
  \end{aligned}
\end{equation}
Taking the inverse gives the action-separation law stated in the main text. If
$\Phi\geq u_1$, then $U_*=u_1/\Phi$ lies in $[0,1]$. When the induced
density is nonzero there, one-sidedly if $U_*=1$, and the zero has finite
order, the same local calculation as in the Loschmidt-zero transition gives a
nonintegrable inverse-Fisher tail, hence infinite CVaR and zero tail-certified
information.

\section{RTN coherence and exact safety threshold}
\label{app:rtn}

For a symmetric RTN process $n(t)=\pm1$ switching at rate $\gamma_{\rm RTN}$, define
$g_\pm(t)=\E[\exp(i\lambda\int_0^t n(s)ds)\mid n(0)=\pm1]$. The sum
$S=g_++g_-$ obeys
\begin{equation}
  \ddot S+2\gamma_{\rm RTN}\dot S+\lambda^2S=0,\qquad S(0)=2,\quad \dot S(0)=0.
  \label{eq:rtn-ode}
\end{equation}
For $\lambda>\gamma_{\rm RTN}$ this gives
\begin{equation}
  \begin{aligned}
  &W_{\rm RTN}(T;\lambda,\gamma_{\rm RTN})\\
  &\quad =
  e^{-\gamma_{\rm RTN} T}
  \biggl[
  \cos(\Omega T)
  +
  \frac{\gamma_{\rm RTN}}{\Omega}\sin(\Omega T)
  \biggr],\\
  \Omega&=\sqrt{\lambda^2-\gamma_{\rm RTN}^2}.
  \end{aligned}
\label{eq:rtn-coherence}
\end{equation}
Zeros solve
\begin{equation}
\begin{gathered}
  \cos(\Omega T)
  +
  \frac{\gamma_{\rm RTN}}{\Omega}\sin(\Omega T)=0,\\
  \Omega=\sqrt{\lambda^2-\gamma_{\rm RTN}^2}.
\end{gathered}
\end{equation}
The first zero is
\begin{equation}
  T_{\rm zero}(\lambda,\gamma_{\rm RTN})
  =
  \frac{\pi/2+\arcsin(\gamma_{\rm RTN}/\lambda)}
  {\sqrt{\lambda^2-\gamma_{\rm RTN}^2}},
\end{equation}
which decreases with $\lambda$. To see monotonicity, write
$\delta=\gamma_{\rm RTN}/\lambda$ and
$T_{\rm zero}(\lambda,\gamma_{\rm RTN})=h(\delta)/\lambda$, with
$h(\delta)=(\pi/2+\arcsin\delta)/\sqrt{1-\delta^2}$. Since $h>0$ and
$h'>0$ for $0\leq\delta<1$,
$dT_{\rm zero}/d\lambda=-(h+\delta h')/\lambda^2<0$. Therefore the strongest
coupling $\Lambda$ sets the first zero entering the support,
\begin{equation}
  T_{\rm safe}(\Lambda,\gamma_{\rm RTN})
  =
  \frac{\pi/2+\arcsin(\gamma_{\rm RTN}/\Lambda)}
  {\sqrt{\Lambda^2-\gamma_{\rm RTN}^2}} .
  \label{eq:tsafe}
\end{equation}
When $\gamma_{\rm RTN}\geq\lambda$, the hyperbolic continuation is a
positive sum of hyperbolic functions and has no real zero; hence the
zero-driven divergence is absent in the motional-narrowing regime
$\gamma_{\rm RTN}\geq\Lambda$.

\section{Ramsey QFI derivation}
\label{app:ramsey-qfi}

At the optimal working point, the Ramsey output state has Bloch vector
length $\mathcal{W}(T,\xi)$ and signal phase speed $T$. The quantum Fisher information
for the phase parameter is therefore $F_Q=T^2\mathcal{W}^2(T,\xi)$. Measuring in the
quadrature basis at the working point saturates this QFI to leading order,
so the same conditional Fisher information appears in the classical readout.

\section{Bernoulli sparse-bath approximation and reserve allocation}
\label{app:bernoulli}

In the low-occupancy approximation $N=0$ with probability $1-N_f$ and $N=1$
with probability $N_f$, the quasi-static scalar loss is
$a=1/[nT^2C_0^2(T)]$ for $N=0$ and $a\sec^2(\lambda T)$ for $N=1$. For
$\Lambda T<\pi/2$ and $\alpha<1-N_f$,
\begin{equation}
  \CVaR_\alpha(1/F_Q)
  =
  a\left[
  1+\frac{N_f}{1-\alpha}
  \left(\frac{\tan(\Lambda T)}{\Lambda T}-1\right)
  \right],
  \label{eq:ramsey-rp-approx}
\end{equation}
This gives the closed-form design expression used for the single-arm
$T_{\rm TC}$ optimization. For a safe reserve anchor, the Poisson
moment-generating function gives~\eqref{eq:reserve-moments}; the first-moment
bound gives the sufficient shot rule
\begin{equation}
  n_0
  \geq
  \frac{
  \exp\!\left[
  N_f(\sec^2(\Lambda T_0)-1)+2(T_0/\Ttwostar)^2
  \right]}
  {(1-\alpha)RT_0^2}.
  \label{eq:reserve-shot}
\end{equation}
Here $R$ is the target upper bound on the reserve contribution to the CVaR
loss.

\section{Proof of the Ramsey portfolio theorem}
\label{app:portfolio-proof}

For $K$ Ramsey arms with positive shot allocations, the quasi-static sparse
Poisson portfolio Fisher information is
\begin{equation}
  F_{\rm port}(\xi)
  =
  \sum_{j=1}^K
  n_jT_j^2C_0^2(T_j)
  \prod_{r=1}^N\cos^2(\lambda_rT_j).
  \label{eq:portfolio-fi}
\end{equation}
For arm $j$, the zero grid inside the coupling support is
\begin{equation}
  \mathcal G_j
  :=
  \left\{\frac{(2k+1)\pi}{2T_j}:k=0,1,2,\ldots\right\}\cap[0,\Lambda].
  \label{eq:zero-grid-new}
\end{equation}

Condition on $N$ active fluctuators. The zero set of
Eq.~\eqref{eq:portfolio-fi} is a finite union of strata on which every arm is
zeroed by at least one coupling coordinate. Zero-nondegeneracy of the grids
$\mathcal G_j$ implies that a single coupling coordinate cannot zero two
different arms. Therefore the generic portfolio-zero strata require $K$
distinct coupling coordinates, one per arm, and have codimension $K$.

Near such a generic stratum choose normal coordinates
$u_j=\lambda_{r_j}-\lambda_j^*$ with $\lambda_j^*\in\mathcal G_j$. Since
$\cos(\lambda T_j)=\pm T_j u_j+O(u_j^2)$ at a simple zero,
\begin{equation}
  F_{\rm port}(\xi)
  =
  \sum_{j=1}^K b_j u_j^2+O(\|u\|^3),
  \qquad b_j>0,
\end{equation}
after fixing all tangential coordinates away from additional zeros. The
volume of the set $\sum_j b_j u_j^2<x^{-1}$ is proportional to
$x^{-K/2}$. The coupling density is bounded above and below by positive
constants at the chosen zeros, so the same exponent holds for the
conditional probability. Strata involving repeated fluctuators, multiple
zeros associated with the same arm, or accidental additional constraints do
not lower the logarithmic tail exponent below $K/2$. Under
zero-nondegeneracy, the leading generic event is one independent simple zero
per arm; nongeneric strata contribute only subleading constants or
logarithmic corrections. Summing over the Poisson distribution of $N$ changes
the leading constant but not the logarithmic exponent. In the sparse limit
the first nonzero generic contribution comes from $N\geq K$ and is
proportional to $N_f^K$.

\section{Proof of the Fisher reserve theorem}
\label{app:reserve-proof}

For the safe anchor, $T_0<T_{\rm safe}$ implies
$m_0=\min_{\lambda\in[0,\Lambda]}|W_{\rm RTN}(T_0;\lambda,\gamma_{\rm RTN})|>0$.
Therefore the anchor contribution obeys
\begin{equation}
  F_0(\xi)
  =
  n_0T_0^2C_0^2(T_0)
  \prod_{k=1}^N W_{\rm RTN}^2(T_0;\lambda_k,\gamma_{\rm RTN})
  \geq a_0m_0^{2N}.
\end{equation}
All other Fisher contributions are nonnegative, hence
$F_{\rm tot}\geq F_0$ and $\ell_{\rm tot}\leq a_0^{-1}m_0^{-2N}$. Raising to
power $s>0$ and using the Poisson moment-generating function gives
\begin{equation}
  \E[\ell_{\rm tot}^s]
  \leq
  a_0^{-s}\E[m_0^{-2sN}]
  =
  a_0^{-s}\exp[N_f(m_0^{-2s}-1)].
  \label{eq:reserve-moments}
\end{equation}
The first moment is finite, hence every CVaR level below one is finite for
the nonnegative loss. The super-polynomial tail follows by comparing
$\ell_{\rm tot}$ to $C\kappa^N$ and applying standard Poisson tail asymptotics.

\section{Hahn echo proof}
\label{app:echo}

The Hahn-echo RTN coherence can be derived by propagating the RTN transfer
matrix through two intervals of length $T/2$ with opposite signs of the
coupling. In the oscillatory regime, let $\tau=T/2$,
$c=\cos(\Omega\tau)$, $s=\sin(\Omega\tau)$, and
$\epsilon=e^{-\gamma_{\rm RTN}\tau}$. The transfer matrix for the first half is
\begin{equation}
  A_{+\lambda}(\tau)=
  \epsilon
  \begin{pmatrix}
  c+i(\lambda/\Omega)s & (\gamma_{\rm RTN}/\Omega)s\\
  (\gamma_{\rm RTN}/\Omega)s & c-i(\lambda/\Omega)s
  \end{pmatrix},
\end{equation}
and $A_{-\lambda}$ is obtained by replacing $\lambda$ by $-\lambda$. With a
uniform initial RTN state,
\begin{equation}
  W_{\rm echo}(T)
  =
  \frac12(1,1)A_{-\lambda}(\tau)A_{+\lambda}(\tau)
  \begin{pmatrix}1\\1\end{pmatrix},
\end{equation}
which evaluates to
\begin{equation}
\begin{aligned}
  &W_{\rm echo}(T;\lambda,\gamma_{\rm RTN})\\
  &\quad =
  e^{-\gamma_{\rm RTN} T}
  \biggl[
  \left(c+\frac{\gamma_{\rm RTN}}{\Omega}s\right)^2
  +
  \frac{\lambda^2}{\Omega^2}
  s^2
  \biggr].
\end{aligned}
  \label{eq:echo-osc-new}
\end{equation}
Replacing $\Omega\to i\Omega_h$, and writing
$c_h=\cosh(\Omega_h\tau)$ and $s_h=\sinh(\Omega_h\tau)$, gives
\begin{equation}
\begin{aligned}
  &W_{\rm echo}(T;\lambda,\gamma_{\rm RTN})\\
  &\quad =
  e^{-\gamma_{\rm RTN} T}
  \biggl[
  \left(c_h+\frac{\gamma_{\rm RTN}}{\Omega_h}s_h\right)^2
  +
  \frac{\lambda^2}{\Omega_h^2}
  s_h^2
  \biggr],
\end{aligned}
  \label{eq:echo-hyp-new}
\end{equation}
with $\Omega_h=\sqrt{\gamma_{\rm RTN}^2-\lambda^2}$. In the oscillatory
formula, the bracket is a sum of two squares; if the sine term vanishes, then
the cosine term is $\pm1$, so the bracket is nonzero. In the hyperbolic
formula the first square is strictly positive. Thus the single-fluctuator
echo coherence has no real zero in the stated model.

\section{Same-axis DC sensing bound}
\label{app:dc-bound}

Let $y(t)=\pm1$ be a pulse toggling function and
$A_y=\int_0^T y(t)\,dt$. In the quasi-static scalar model,
$W_y(\lambda)=\cos(\lambda A_y)$ and
$F_Q^{(y)}(\lambda)=A_y^2\cos^2(\lambda A_y)$. Avoiding coherence zeros for
all $\lambda\in[0,\Lambda]$ requires $|A_y|<\pi/(2\Lambda)$, which implies
$F_Q^{(y)}(\lambda)<\pi^2/(4\Lambda^2)$. If this condition is violated and
the coupling density is nonzero at the first zero, the simple-zero
$x^{-1/2}$ inverse-Fisher tail follows from the Fisher-codimension law.

\section{Finite-sample simulation details}
\label{app:finite-sample}

\begin{figure*}[tbp]
\centering
\includegraphics[width=\textwidth]{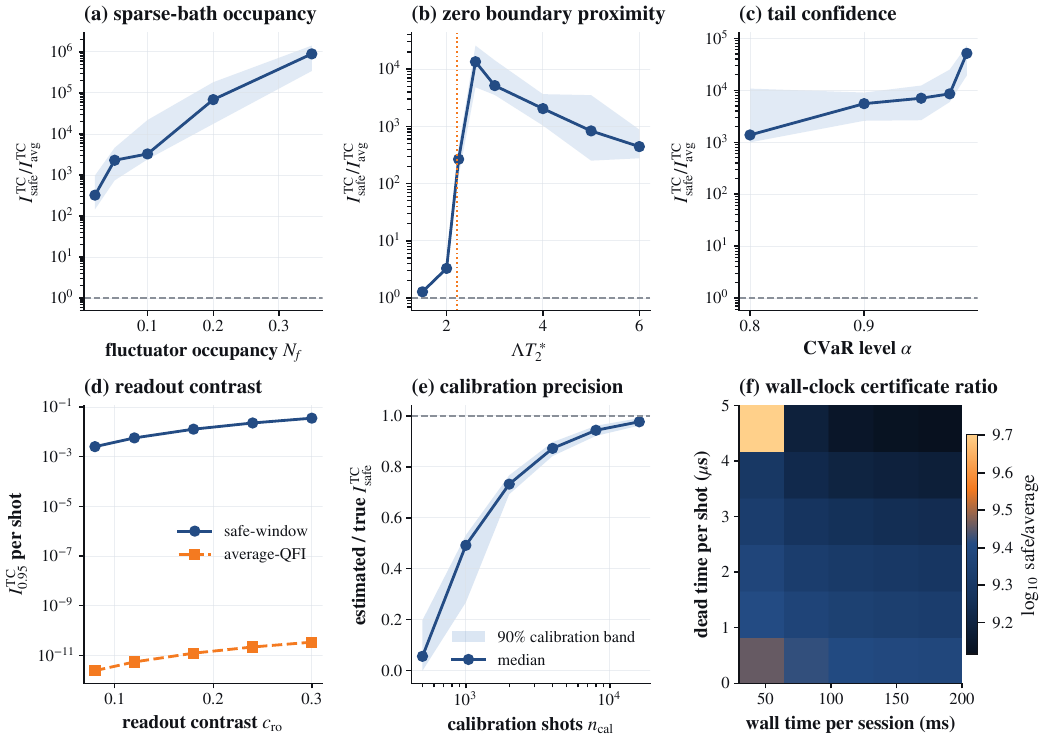}
\caption{\textbf{Sensitivity of the platform-level certificate.}
Sweeps around the representative shallow-NV parameters in
Table~\ref{tab:nv-platform-parameters}. (a)--(c) The finite-session ratio of
safe-window to average-QFI tail-certified information is shown as the median
over $24$ independent latent ensembles with $N_{\rm sess}=32000$ sessions each.
Shading gives a $90\%$ bootstrap confidence interval for the median log-ratio;
the raw $10$--$90\%$ ensemble spread is reported in the processed results file.
The panels vary fluctuator occupancy $N_f$, dimensionless coupling range
$\Lambda T_2^\ast$, and CVaR level $\alpha$. The dotted line in (b) marks
$\Lambda T_2^\ast=\pi/\sqrt{2}$, where the leading average-QFI optimum first
crosses the quasi-static safety boundary. (d) Readout contrast scales the
absolute certificates but does not remove the ranking reversal. (e) Finite
calibration shots bias the empirical safe-window certificate low at small
$n_{\rm cal}$ and converge toward the true session-resolved certificate as
the calibration block grows. (f) With a fixed wall-clock budget per session,
including calibration overhead and dead time per shot, the safe-window
advantage persists over the swept time budgets.}
\label{fig:nv-sensitivity-sweeps}
\end{figure*}

\paragraph*{Supplementary platform robustness figures.}
Figure~\ref{fig:nv-sensitivity-sweeps} varies the main experimental knobs
around Table~\ref{tab:nv-platform-parameters}. The distinction is controlled
most sharply by $\Lambda T_2^\ast$: below the zero-crossing threshold the
average-QFI optimum remains inside the safe window and the ranking difference
is modest, while above it the average-QFI tail certificate collapses and the
safe-window certificate remains strongly favorable across the swept density,
CVaR level, readout contrast, calibration-shot, and wall-clock settings.

\begin{figure*}[tbp]
\centering
\includegraphics[width=\textwidth]{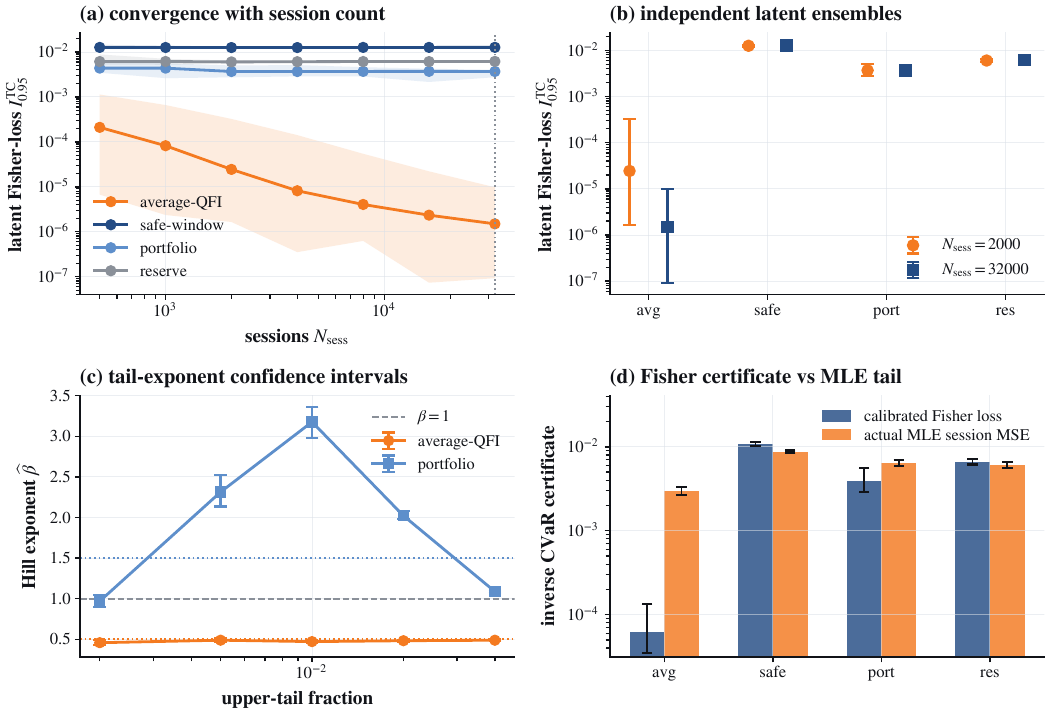}
\caption{\textbf{Finite-sample robustness of the shallow-NV certificate.}
(a) Convergence of the latent Fisher-loss tail-certified information with
session count $S$ over $24$ independent latent ensembles; bands show
$10$--$90\%$ seed-to-seed variation and the dotted line marks the high-stat
tournament size $N_{\rm sess}=32000$. (b) Independent-ensemble uncertainty at
$N_{\rm sess}=2000$ and $N_{\rm sess}=32000$.
(c) Fixed-threshold bootstrap confidence intervals for the Hill upper-tail
exponent. The average-QFI arm lies below the finite-CVaR boundary
$\beta=1$, while the portfolio steepens the finite-tail diagnostic. (d)
Direct comparison between calibrated Fisher-loss inverse CVaR and actual
MLE session mean-square-error (MSE) inverse CVaR, using $5000$ sessions and
$R_{\rm rob}=16$ sensing repeats per session. Error bars in (d) are
session-bootstrap $90\%$ intervals.}
\label{fig:nv-robustness-checks}
\end{figure*}

Figure~\ref{fig:nv-robustness-checks} checks finite-sample robustness.
Increasing the number of latent sessions stabilizes the certified designs
while the unsafe average-QFI arm continues to sample rarer near-zero
sessions. Hill intervals place the unsafe arm below the $\beta=1$
integrability boundary, and the direct MLE-error comparison shows that
calibrated Fisher-loss CVaR tracks the actual estimator ranking once
finite-estimator regularization is included.

\paragraph*{Platform-level simulation algorithm.}
The shallow-NV platform demonstration uses the following measurement-level
simulation. The input parameters are
\begin{align*}
  &N_{\rm sess},\ n,\ n_{\rm cal},\ T_2^\ast,\ c_{\rm ro},\ N_f,\ \Lambda,\ \theta_0,\\
  &\mathcal T=\{(T_k,w_k)\}.
\end{align*}
The executable implementation is split into a slow experiment/cache script and
a fast figure-rendering script. Running
\texttt{python numerics/tail\_certification\_experiments.py --figure all}
reruns the manuscript simulations, writes fixed numerical caches to
\url{numerics/output/data/}, and writes processed summaries to
\url{numerics/output/results.json} and
\url{numerics/output/data/plot_cache_manifest.json}. Style-only figure
changes can then be rendered without rerunning the experiments by
\texttt{python numerics/tail\_certification\_figures.py --figure all}. All
pseudorandom numbers use NumPy \texttt{default\_rng} with base seed
$20260430$.

The one-arm times are selected by the same finite design tournament used in
the plotted baseline. A separate \emph{design} ensemble of
$N_{\rm design}=2600$ latent configurations is drawn with the
Fig.~\ref{fig:nv-platform-demo} stream, distinct from the $N_{\rm sess}=32000$
sessions used to evaluate the tournament. The grid
$T\in[0.12,4.6]\,\mu{\rm s}$ has $260$ equally spaced points. The average-QFI
time $T_{\rm avg}$ maximizes
$N_{\rm design}^{-1}\sum_{i=1}^{N_{\rm design}}(c_{\rm ro} C_i(T)T)^2$ on this
grid. The CVaR time $T_{\rm TC}$ maximizes
$\widehat{\mathcal{I}}^{\rm TC}_{0.95}(T)=
1/\widehat{\rm CVaR}_{0.95}\{[c_{\rm ro}C_i(T)T]^{-2}\}$ over
$T<T_{\rm safe}=\pi/(2\Lambda)$, with the certificate set to zero outside the
safe window. This gives
$T_{\rm avg}=2.109189\,\mu{\rm s}$ and
$T_{\rm TC}=0.950270\,\mu{\rm s}$ for the Table~\ref{tab:nv-platform-parameters}
baseline.

The four fixed tournament protocols are: average-QFI,
$(T,w)=(T_{\rm avg},1)$; safe-window,
$(T,w)=(T_{\rm TC},1)$; portfolio,
$T=(1.85,2.20,2.65)\,\mu{\rm s}$ with equal weights; and reserve,
$T=(1.00,T_{\rm avg})\,\mu{\rm s}$ with weights $(0.30,0.70)$. For
$n=2000$ sensing shots the integer shot allocations are respectively
$2000$, $2000$, $(666,666,668)$, and $(600,1400)$; integer counts are formed
by flooring $nw_k$ and assigning the residual shots to the last arm.
For each session $i=1,\ldots,N_{\rm sess}$:
\begin{enumerate}
  \item Draw $N_i\sim{\rm Poisson}(N_f)$.
  \item Draw latent couplings $\lambda_{ij}\sim{\rm Uniform}[0,\Lambda]$.
  \item Compute
  $C_i(T_k)=\exp[-(T_k/T_2^\ast)^2]\prod_j\cos(\lambda_{ij}T_k)$.
  \item Generate calibration shots at phases $\varphi=0,\pi$. For a protocol
  with $K$ interrogation times, use
  $n_{\rm phase}=\max\{20,\lfloor n_{\rm cal}/(2K)\rfloor\}$ shots for each
  phase at each time.
  \item Estimate
  $\widehat C_i(T_k)=(Y_{ik}^+/n_{\rm phase}-Y_{ik}^-/n_{\rm phase})/c_{\rm ro}$,
  clipped to $[-1,1]$, and
  $\widehat F_i=\sum_k w_k[c_{\rm ro}\widehat C_i(T_k)T_k]^2$.
  \item Generate sensing shots at $\varphi=\pi/2$ using the integer shot
  allocation above.
  \item Estimate $\widehat\theta_i$ by a one-dimensional bounded Newton MLE.
  The search starts at $\theta_0$, uses at most $12$ Newton steps, clips
  Bernoulli probabilities to $[10^{-7},1-10^{-7}]$, and constrains the iterate
  to
  $|\theta-\theta_0|\leq
  \min\{0.70,0.45\pi/\max_k T_k\}$. The update stops when the Newton step is
  below $10^{-8}$ or when the Hessian is nonfinite or smaller than $10^{-12}$
  in magnitude. No nuisance coordinate is fitted in the reported NV panels.
  \item Store $\widehat\ell_i=\widehat F_{i,\theta|\mu}^{-1}$ and
  $(\widehat\theta_i-\theta_0)^2$.
\end{enumerate}
The calibration shot counts are therefore exact for the single-arm and
reserve designs; for the three-arm portfolio the calibration uses
$3996$ shots because
$2K\lfloor4000/(2K)\rfloor=3996$.

The seed protocol is deterministic. Fig.~\ref{fig:nv-platform-demo} uses
stream $20260430+301$ for latent sessions and calibration, and independent
MLE-error streams $20260430+9000+p$ for protocol index
$p=0,1,2,3$ in the order average-QFI, safe-window, portfolio, reserve.
Fig.~\ref{fig:nv-sensitivity-sweeps} uses streams
$20260430+1400+100j+r$ for the $N_f$ sweep,
$20260430+1500+100j+r$ for the coupling-range sweep,
$20260430+1600+100j+r$ for the CVaR-level sweep,
$20260430+1700$ and $20260430+1701$ for the baseline/readout-contrast
calculation, $20260430+1800$ for calibration latent configurations, and
$20260430+1900+100j+r$ for calibration resampling. Here $j$ indexes the
swept value and $r=0,\ldots,23$ indexes the replicate for panels (a)--(c).
Fig.~\ref{fig:nv-robustness-checks}
uses streams $20260430+22000+s$ for the $24$ independent convergence
ensembles, $20260430+24000$ for Hill latent sessions,
$20260430+24500+100j+\delta$ for Hill bootstraps with
$\delta=0$ for average-QFI and $50$ for portfolio, $20260430+25000$ for the
MLE-validation latent sessions, $20260430+26000+p$ for MLE sensing repeats,
and $20260430+27000+p$ and $20260430+28000+p$ for Fisher-loss and MLE-error
bootstrap intervals.

The outputs are the latent average Fisher information, the empirical
$\widehat{\mathcal{I}}^{\rm TC}_{\alpha}$, the empirical
$\widehat{\rm CVaR}_\alpha[(\widehat\theta-\theta_0)^2]$, the Hill estimate
$\widehat\beta(u)$, and the local attainability ratio
$n\,\widehat{\rm Var}(\widehat\theta)\widehat F$. The reported
tail-certified intervals are obtained by nonparametric session-bootstrap
resampling of the calibrated losses. The MLE-error survival in
Fig.~\ref{fig:nv-platform-demo} pools $R_{\rm err}$ repeated sensing blocks
per calibrated session. The attainability ratio is estimated by repeating the
sensing block at fixed calibrated latent sessions, matching Eq.~\eqref{eq:nv-attainability-ratio}.

For the quasi-static sparse Ramsey model
\begin{equation}
  \begin{aligned}
  F_Q(T,\xi)
  &=
  T^2C_0^2(T)\prod_{k=1}^{N}\cos^2(\lambda_kT),\\
  \lambda_k&\sim{\rm Uniform}[0,1],
  \end{aligned}
\end{equation}
with $N_f=0.05$, $\alpha=0.9$, and $T_2^\star=5$, the average-QFI optimum is
$T_{\rm avg}=3.5355$, while $T_{\rm safe}=\pi/2=1.5708$ and the Bernoulli design formula
gives $T_{\rm TC}=1.2030$. Controlled tail-estimation tests recover the
$x^{-1/2}$ simple-zero slope, the codimension law $c/2$ for synthetic Fisher
nulls, and the positive-channel cut exponent $K/2$. For a full Poisson bath
with $N_f=1$, adding a safe anchor at $T_0=1.2<T_{\rm safe}$ reduces representative
extreme-loss quantiles by orders of magnitude, illustrating reserve-induced
removal of the algebraic tail.

The sensitivity sweeps in Fig.~\ref{fig:nv-sensitivity-sweeps} use the same
session-resolved latent model as Fig.~\ref{fig:nv-platform-demo}, but
evaluate the calibrated Fisher-loss certificate rather than rerunning the
full MLE-error survival for every point. Panels (a)--(c) use
$24$ independent latent ensembles with $N_{\rm sess}=32000$ sessions each.
They plot the median safe-window/average-QFI ratio with a $90\%$ bootstrap
confidence interval for the median log-ratio; the corresponding raw
$10$--$90\%$ spread across finite-session ensembles is retained in
\texttt{numerics/output/results.json}. Panel (d) varies $c_{\rm ro}$ at fixed
latent ensembles.
Panel (e) repeats the two-phase calibration shot model and reports the
ratio between the calibration-estimated and true safe-window certificate.
Panel (f) converts per-shot certificates to fixed-wall-time certificates by
using
\[
  n_{\rm eff}
  =
  \frac{T_{\rm wall}
  -
  n_{\rm cal}\langle T_k+t_{\rm dead}\rangle_{\rm cal}}
  {\langle T_k+t_{\rm dead}\rangle_{\rm sense}},
\]
with negative values clipped to zero, and then multiplying the per-shot
tail-certified information by $n_{\rm eff}$. Here $T_{\rm wall}$ is the
per-session wall-clock budget and $t_{\rm dead}$ is the per-shot dead time.

The robustness checks in Fig.~\ref{fig:nv-robustness-checks} use the
baseline parameters in Table~\ref{tab:nv-platform-parameters}. For panels
(a) and (b), we draw $24$ independent latent ensembles of size $32000$ and
evaluate the true latent Fisher-loss CVaR on prefixes
$N_{\rm sess}=500,1000,2000,4000,8000,16000,32000$. This isolates sampling convergence
from finite calibration-shot zeros; the calibrated certificate is used again
in the estimator comparison. At the tournament size $N_{\rm sess}=32000$ and
$\alpha=0.95$, the empirical CVaR averages the largest $1600$ losses; the
smaller $N_{\rm sess}=2000$ prefix averages only $100$ losses and is retained as a
finite-session diagnostic. Panel (c) estimates Hill exponents from
$240000$ latent sessions and reports fixed-threshold bootstrap $90\%$
intervals for upper-tail fractions $0.002$--$0.04$. Panel (d) draws $5000$
calibrated latent sessions, runs $R_{\rm rob}=16$ independent MLE sensing
repeats per session, and compares $\widehat{\rm CVaR}_{0.95}(\widehat F^{-1})$
with $\widehat{\rm CVaR}_{0.95}[n\,R_{\rm rob}^{-1}\sum_r
(\widehat\theta_{ir}-\theta_0)^2]$ over sessions. The error bars are
nonparametric session-bootstrap $90\%$ intervals.

\section{Regulator-flow proof}
\label{app:regulator-flow}

Let $F\geq0$ be the session-resolved projected Fisher information and assume
\begin{equation}
  \Prob(F<f)=A_\beta f^\beta+o(f^\beta),
  \qquad f\downarrow0 .
  \label{eq:app-reg-small-f}
\end{equation}
Then $\ell=F^{-1}$ satisfies
\begin{equation}
  \Prob(\ell>x)=A_\beta x^{-\beta}+o(x^{-\beta}).
  \label{eq:app-reg-loss-tail}
\end{equation}

For a capped loss $\ell^{\rm cap}=\min(\ell,\ell_{\max})$, the upper-tail
representation of CVaR gives, for fixed $\alpha<1$ and $\ell_{\max}$ above
the $\alpha$-quantile $q_\alpha$,
\begin{equation}
  \CVaR_\alpha(\ell^{\rm cap})
  =
  q_\alpha+
  \frac{1}{1-\alpha}
  \int_{q_\alpha}^{\ell_{\max}}\Prob(\ell>x)\,dx .
  \label{eq:app-cap-cvar-tail}
\end{equation}
The finite term $q_\alpha$ does not affect the regulator divergence.
Therefore, for $0<\beta<1$,
\begin{equation}
  \CVaR_\alpha(\ell^{\rm cap})
  =
  O(1)+
  \frac{A_\beta}{(1-\alpha)(1-\beta)}\ell_{\max}^{1-\beta},
  \label{eq:app-cap-beta-lt-one}
\end{equation}
while for $\beta=1$,
\begin{equation}
  \CVaR_\alpha(\ell^{\rm cap})
  =
  O(1)+
  \frac{A_\beta}{1-\alpha}\log \ell_{\max} .
  \label{eq:app-cap-beta-one}
\end{equation}
This proves the hard-cap scaling.

For an additive Fisher floor $\ell_\varepsilon=(F+\varepsilon)^{-1}$, write the
lower-tail measure as
\begin{equation}
  d\Prob(F<f)=A_\beta\beta f^{\beta-1}df+o(f^{\beta-1})\,df .
  \label{eq:app-lower-tail-density}
\end{equation}
The divergent part of the upper-tail conditional expectation is
\begin{equation}
  \frac{1}{1-\alpha}
  \int_0^{f_\alpha}
  \frac{A_\beta\beta f^{\beta-1}}{f+\varepsilon}\,df .
  \label{eq:app-floor-integral}
\end{equation}
For $0<\beta<1$, set $f=\varepsilon u$. Then
\begin{equation}
  \int_0^{f_\alpha}
  \frac{f^{\beta-1}}{f+\varepsilon}\,df
  =
  \varepsilon^{\beta-1}
  \int_0^{f_\alpha/\varepsilon}
  \frac{u^{\beta-1}}{1+u}\,du
  \sim
  \varepsilon^{\beta-1}
  \frac{\pi}{\sin(\pi\beta)} .
  \label{eq:app-floor-beta-integral}
\end{equation}
Hence
\begin{equation}
  \CVaR_\alpha(\ell_\varepsilon)
  \sim
  \frac{A_\beta\beta\pi}{(1-\alpha)\sin(\pi\beta)}
  \varepsilon^{\beta-1}.
  \label{eq:app-floor-cvar-beta-lt-one}
\end{equation}
For $\beta=1$, the same integral gives
\begin{equation}
  \CVaR_\alpha(\ell_\varepsilon)
  \sim
  \frac{A_\beta}{1-\alpha}\log(1/\varepsilon).
  \label{eq:app-floor-cvar-beta-one}
\end{equation}

If $F=A f$, then
\begin{equation}
  (F+\varepsilon)^{-1}
  =
  A^{-1}(f+\varepsilon/A)^{-1},
  \label{eq:app-amplified-floor}
\end{equation}
so
\begin{equation}
  \mathcal{I}^{\rm TC}_{\alpha,\varepsilon}(A)
  \propto
  A^\beta\varepsilon^{1-\beta}
  \label{eq:app-amplified-regulated-scaling}
\end{equation}
for $0<\beta<1$. Thus a nonintegrable Fisher-zero tail changes the regulated
scaling exponent from $A$ to $A^\beta$.

For a finite empirical latent catalog $f_1,\ldots,f_N$ drawn from
$\Prob(f<y)\sim C y^\beta$, let $\ell_{(k)}^\downarrow$ be the $k$th largest
inverse loss. Extreme-value order statistics give
\begin{equation}
  \ell_{(k)}^\downarrow
  \asymp
  A^{-1}\left(\frac{CN}{k}\right)^{1/\beta}.
  \label{eq:app-order-stat}
\end{equation}
Averaging the largest $(1-\alpha)N$ losses gives
\begin{equation}
  \widehat{\CVaR}_\alpha(\ell)
  \asymp
  A^{-1}N^{1/\beta-1},
  \qquad 0<\beta<1,
  \label{eq:app-catalog-cvar-beta-lt-one}
\end{equation}
and therefore
\begin{equation}
  \widehat{\mathcal{I}}^{\rm TC}_{\alpha,N}
  \asymp
  A N^{-(1-\beta)/\beta}.
  \label{eq:app-catalog-itc-beta-lt-one}
\end{equation}
At $\beta=1$,
\begin{equation}
  \widehat{\CVaR}_\alpha(\ell)
  \asymp
  A^{-1}\log N,
  \qquad
  \widehat{\mathcal{I}}^{\rm TC}_{\alpha,N}
  \asymp
  \frac{A}{\log N}.
  \label{eq:app-catalog-beta-one}
\end{equation}
Finite latent catalogs therefore round, but do not remove, the Fisher-glass
scaling.

\section*{Data and code availability}

The simulation code and cached data that reproduce every main-text and
supplemental figure, and every reported number, are openly available at
\url{https://github.com/E-zClap/fisher_glass_data}. All results are
simulation-based; no unpublished experimental data are used, and the analytic
derivations are contained in the appendices. The numerical routines require only
\texttt{numpy} and \texttt{matplotlib}: the shared library and figure driver is
\texttt{tail\_certification\_numerics.py}, and the conceptual figures are built
by \texttt{quenched\_fisher\_figures.py}. Under a fixed random seed, each figure
can be redrawn from the shipped caches without rerunning the simulations
(\texttt{--figures-only}) or recomputed from scratch (\texttt{--refresh-data});
exact commands are given in the repository README.

\end{document}